# Impact of power outages on the adoption of residential solar photovoltaic in a changing climate

Jiashu Zhu[1*], Wenbin Zhou[2], Laura Diaz Anadon[1], Shixiang Zhu[2*]

[1] Cambridge Centre for Environment, Energy and Natural Resource Governance, Department of Land Economy, University of Cambridge, United Kingdom.

[2] Heinz College of Information Systems and Public Policy, Carnegie Mellon University, United States.

* Corresponding authors: Jiashu Zhu (jz568@cam.ac.uk), Shixiang Zhu (shixianz@andrew.cmu.edu)

**Abstract**

Residential solar photovoltaic (PV) systems are a cornerstone of residential decarbonization and energy resilience. However, most existing systems are PV-only and cannot provide backup power during grid failures. Here, we present a high-resolution analysis of 377,726 households in Indianapolis, US, quantifying how power outages influence the installation of PV-only systems between 2014 and 2023. Using a two-part econometric panel model, we estimate the causal effect of power outage exposure and project future risks under a middle of the road climate scenario (RCP 4.5). We find that each additional hour of annual outage duration per household lowers the new-installation rate by 0.012 percentage points per year, equivalent to a 31% decline relative to the historical mean (2014-2023). With outage duration and frequency projected to double by 2040, these results reveal a potential vicious cycle between grid unreliability and slower decarbonization, calling for policies that integrate grid resilience and clean-energy goals.

# 1    Introduction

Electricity grids are a critical enabler of clean energy transitions, connecting the growing supply of renewable power to rising demand from electrification. As grids continue to expand, their reliability is frequently challenged by aging infrastructure and extreme weather events[1–4]. Large-scale and cascading blackouts in regions such as California, Chile, and the Iberian Peninsula have led to growing concerns about energy resilience among policymakers and the public[5,6]. Studies from developing countries demonstrate a strong correlation between exposure to power outages and the uptake of off-grid solar PV, as households seek to compensate for unreliable utility service[7,8]. In contrast, residential solar PV in developed countries has typically been grid-connected, with adoption driven primarily by economic incentives[6,9]. However, these incentives have been scaled back in parts of the US and Europe as policy support shifts from net metering, under which households receive a full retail-rate credit for any excess electricity they send to the grid, to net billing tariffs in which exported generation is credited at a much lower wholesale-based price[10,11]. This shift reduces savings from the adoption of residential solar PV for households without battery storage or flexible demand to self-consume their generation[12]. While net billing can incentivize the adoption of battery storage and reduce cross-subsidies, it lowers the financial returns and extends the payback period of PV-only systems[12–16].

Given this evolving policy landscape and rising exposure to extreme weather, understanding how exposure to power outages influences the adoption of distributed energy resources (DERs) becomes increasingly relevant. A recent survey of DER adopters in the US found that providing backup power and achieving energy independence was the second most cited motivation, with 72% of households identifying it as a primary driver[17]. Beyond providing back-up power for households, residential solar PV and batteries also support the wider electricity system. Regions transitioning away from aging thermal generation fleets, such as the Midwestern US and parts of Europe, face growing risks of load shedding and generation curtailment[18,19]. Load shedding occurs when heatwaves or cold spells push electricity demand beyond generation capacity, requiring utilities to disconnect some customers to prevent a system-wide collapse[20]. At the same time, renewable power is frequently curtailed because congested or under-sized transmission lines cannot carry power from generation-rich areas to where it is needed most[21,22]. Residential solar PV offers a quickly deployable and cost-effective solution by producing power where it is used, reducing peak demand and therefore lowering reliance on centralized power generation and transmission assets[23,24]. A deeper understanding of the effect of power outages on PV adoption offers new insights necessary for guiding forward-looking grid investments and improving forecasts of technology adoption that account for the effects of outages, as well as for designing policies to promote the adoption of DERs.

The importance of obtaining these insights is growing since extreme weather events, a main driver of power outages, are becoming more frequent and severe under climate change[25–29]. In the US, extreme weather caused around half of all power outages and was responsible for 83% of the customers who experienced outages between 2003 and 2017[27]. The US average annual duration of power interruptions per customer has surged by 181%, from 236 minutes in 2014 to 662 minutes in 2024[30]. Globally, rising temperatures are projected to increase the energy demand by 11-27% by 2050 under the Representative Concentration Pathway (RCP) 4.5 scenario, which is an intermediate climate scenario consistent with a 50% probability of limiting warming to 3°C[31]. This highlights the need for greater integration of DERs into grid planning to ensure clean, reliable, and affordable power supply[32].

Previous research has examined the impact of power outages on the adoption of several energy technologies, including battery storage, PV-plus-storage systems, and electric vehicles[33–37]. While these studies have advanced understanding of outage-driven energy technology adoption, many of them focus on California's Public Safety Power Shutoffs[33,35,37], which are pre-announced power shutoffs implemented in a region with the highest DER penetration rate in the US and a distinctive socioeconomic context, limiting their broader applicability. Recent studies extend the analysis to the national scale but rely on aggregated city- or county-level outage data due to the limited availability of finer-grained datasets[34,36]. Such aggregation can obscure within-city and inter-household heterogeneity in outage exposure arising from local differences in infrastructure quality, maintenance practices, or vegetation cover[17]. In practice, this means that these studies treat energy-technology adopters unaffected by power outages as if they had experienced them (for example, those occurring in remote industrial zones), which can lead to imprecise measurement of treatment exposure and potential bias in the estimated effects.

Despite growing interest in the topic, the impact of power outages on PV-only systems in developed countries has received little attention, most likely due to difficulties in obtaining fine-grained data on power outages. This gap is particularly important given the scale of deployment of such systems. Developed countries account for a substantial share of the global residential solar PV market. Without considering other advanced economies, the US and Europe together had reached a cumulative installed capacity of residential PV of 50 gigawatts (GW) by 2023, representing nearly one-third of the global total[39,40]. Although battery storage deployment has expanded rapidly in recent years[6], PV-only systems still represent the majority of existing installations. By 2023, global installed capacity of residential batteries reached 15 GW, equivalent to 8% of total residential solar capacity[41,42]. The share is slightly higher in developed countries, for example, 14% of US residential solar PV adopters owned storage in 2023, leaving a large stock of PV-only systems[43].

To address these gaps, this study presents a high-resolution assessment of 377,726 households, covering around 890,000 residents in Indianapolis, US, estimating the causal effect of power outages on the adoption of residential PV-only systems. Our comprehensive dataset comprises 91,763 node-level outage events and 1,483 household-level installation records across 44 substations from January 2014 through December 2023. We augment these fine-grained and normally unavailable datasets with hourly 10-meter wind gust speed data at 3-kilometer resolution from the High-Resolution Rapid Refresh (HRRR) atmospheric model[44,45]. Using these datasets, we estimate a two-part econometric panel model using wind gust speed as an instrumental variable (IV). To assess how the impact of outages on adoption may evolve under climate change, we produce probabilistic forecasts of outage risks to 2040 using a hierarchical probabilistic conformal prediction (HPCP) model applied to downscaled global climate model (GCM) data[46,47], under a plausible greenhouse gas concentrations scenario (RCP 4.5). Under this scenario, average temperature in Indianapolis is projected to rise by 1.8°C by 2040, relative to the 1915-2013 baseline[47].

Our findings show that in developed countries, where most residential solar PV systems are grid-connected, exposure to longer power outages reduces the willingness to adopt PV-only systems. Specifically, we find that a one-hour increase in annual power outage duration per household reduces the annual new-installation rate by 0.012% in absolute terms. The annual new-installation rate is the share of newly installed PV-only systems in a given year divided by the total number of households in the study area. Given the low baseline new-installation rate of 0.039% (2014-2023), the causal impact of power outages represents a substantial 31% reduction relative to a counterfactual scenario without outages. We also find that outages exceeding one hour have an even stronger impact, suggesting heightened household sensitivity to major service disruptions, as one may expect. In contrast, outage frequency exhibits weaker and less consistent effects, indicating that shorter but more frequent outages do not trigger the same behavioural responses. The deterrent effect of outages is also strongest in the first 3 months after an outage and diminishes over time, suggesting that the memory of disruptions gradually fades. Our projections under the RCP 4.5 scenario indicate that, all else being equal, both outage duration and frequency are expected to double by 2040 relative to the 2014-2023 average, indicating that without grid upgrades or policy support to foster PV-plus-storage adoption, deterrence on solar PV adoption due to outages would increase even further.

## 2 Results

**Assessing within-city heterogeneity of power outages and solar PV adoption**

We first describe quantitatively the evolution of power outages, weather events, and adoption of PV-only systems over the study period. Indianapolis is a predominantly urban metropolitan area served by a single investor-owned utility, a structure common across many US cities[48]. It would therefore be expected to exhibit relatively uniform grid reliability compared to regions such as California, where service territories span both urban and rural regions. However, our results reveal substantial within-city heterogeneity in both power outage exposure and solar PV adoption, suggesting that within a single service territory, localized differences may drive significant variation in reliability and technology uptake. These results are likely to be relevant for many other urban areas served by similar utility structures.

Grid reliability is measured using the System Average Interruption Duration Index (SAIDI) and the System Average Interruption Frequency Index (SAIFI), both aggregated at the substation level. A substation is a neighbourhood-level facility that transforms high-voltage electricity into lower-voltage levels suitable for local distribution. In Indianapolis, each substation serves roughly 8,600 households, providing a granular unit for analysing reliability. Both indices include major event days and are computed from utility outage records across 44 substations covering the entire residential area of Indianapolis (see Methods). SAIDI represents the total minutes of electric interruptions the average customer experienced in a given period, while SAIFI is the total numbers of interruptions. Monthly SAIDI (Fig. 1a) and SAIFI (Fig. 1b) reveal strong seasonal patterns and overall upward trends in both outage frequency and duration. The 5$^{th}$-95$^{th}$ percentile ranges of SAIDI and SAIFI across substations have widened by 95% and 67%, respectively, over the 10-year period, suggesting growing spatial heterogeneity in grid reliability. In July 2023, the month with the highest disparity, substations in the top 5% each experienced an average of 81 minutes of total outages, more than 26 times longer than those in the bottom 5%. The distributions of both indices follow a long-tail pattern. Most substations report relatively low outage levels in most months, while a small number experience severe disruptions (Fig. S1).

Wind hazards, untrimmed vegetation, and overhead power lines are strongly correlated with increased power outages, posing a compounding risk to electricity distribution networks[49–51]. Monthly maximum wind gust data at 3-kilometer spatial resolution reveal clear within-city variation in wind exposure, reflecting differences in local terrain (Fig. 1c). For example, the Lafayette Road substation, which is exposed near a reservoir and elevated parkland with limited wind shielding, consistently records higher speeds. Between 2014 and 2023, Indianapolis recorded 302 extreme weather events according to the National Oceanic and Atmospheric Administration's (NOAA) Storm Events Base, with over 47% classified as high wind events or thunderstorms[52]. High winds can topple trees and branches onto overhead lines, causing short circuits, equipment damage, or complete line failures. Over the study period, 45,813 outage events (64% of the total) were attributed to tree interference, overhead-equipment failures, and weather-related causes. These patterns are consistent with broader regional trends. In the Midwestern US, the area affected by thunderstorm winds has expanded nearly fivefold since the 1980s, and the storm intensity is also projected to increase by 13% per degree of warming[53]. Thus, insights from Indianapolis are likely to be generalizable to other Midwestern US cities, where electricity distribution is similarly managed by centralized investor-owned utilities and characterized by comparable exposure to weather-related outage risks.

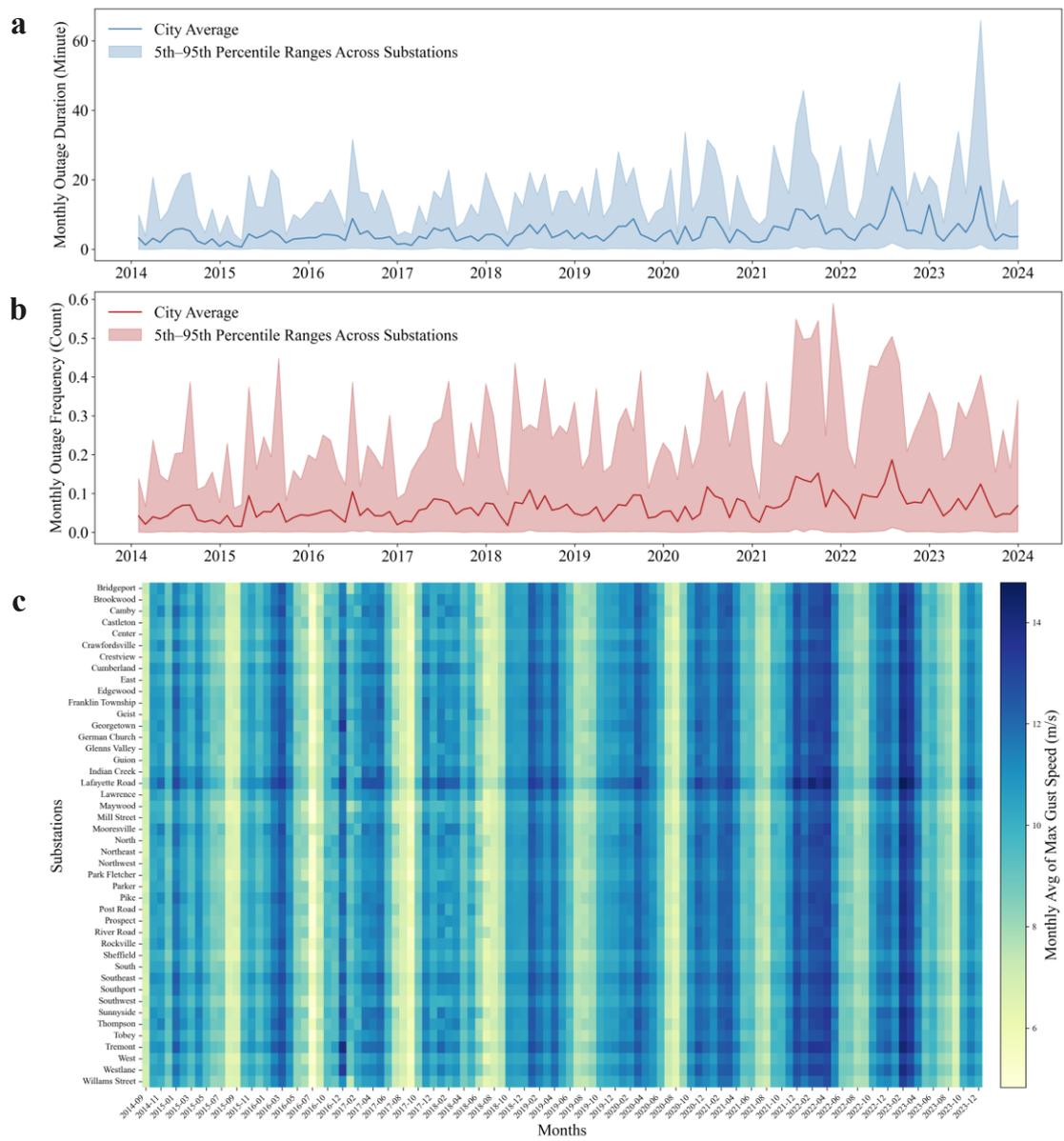

**Fig. 1 | Power outage events and wind gust speeds across substations in Indianapolis. a**, Line plot shows the monthly city-level SAIDI with 5$^{th}$-95$^{th}$ percentile ranges across 44 substations, January 2014-December 2023. **b**, Line plot shows the monthly city-level SAIFI with 5$^{th}$-95$^{th}$ percentile ranges across substations, January 2014-December 2023. Both SAIDI and SAIFI metrics are computed in accordance with the IEEE-1366 standard, including major event days (see Methods). **c**, Heatmap shows the

distribution of maximum monthly wind gust speeds at 10 meters above ground across substations in Indianapolis. The wind gust speed data is extracted from the National Oceanic and Atmospheric Administration's HRRR atmospheric model (see Methods). Wind gust data prior to October 2014 is unavailable as the model only became operational after September 30, 2014[44].

Patterns of residential solar PV adoption reflect similar localized dynamics observed in grid reliability and weather exposure. Citywide residential PV installations accelerated after 2019, reaching 1,483 systems by 2023 (Fig. 2a). The uptake of PV in Indianapolis has been slower than in leading states such as California, where adoption began earlier and accelerated more rapidly[33]. PV and battery storage co-adoption was largely unavailable before 2018 in Indianapolis. About 91% of installed systems in Indianapolis are PV-only, leaving a PV-storage adoption rate of 9% in the city, reflecting the late market entry of battery storage. In contrast, the co-adoption rate had already reached about 14% in California by 2021[34]. Beneath the aggregated trend, adoption trajectories diverge significantly across substations (Fig. 2b). As previously noted, each substation in Indianapolis serves around 8,600 households on average. By 2023, the average adoption rate of PV-only systems had risen above 0.5%. However, adoption remained highly uneven across substations, ranging from fewer than 0.25% to more than 2.5% systems. The divergence widened particularly after 2020, with some substations showing acceleration while others remained relatively flat.

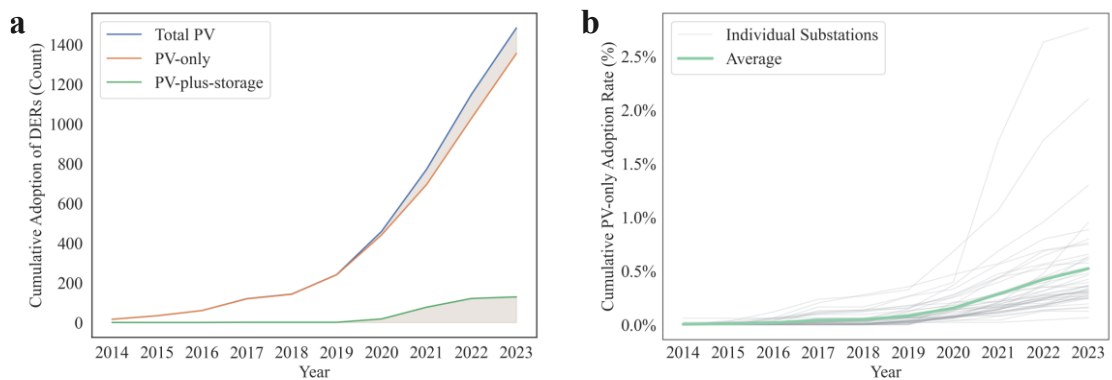

**Fig. 2 | Trend of cumulative adoption of DER systems 2014-2023. a,** Line plot shows the trends of DER installations by year between 2014 and 2023. The blue line is the total number of installed solar PV systems. The orange line represents the total number of PV-only systems. The green line is the number of PV-plus-storage systems. **b,** Line plot shows the cumulative adoption rates of PV-only systems by substation, normalized by the number of customers each substation serves. Each grey line represents the adoption rate for one individual substation. The green line shows the cross-substation average.

Aggregating outage duration, outage frequency, and PV system adoption over the 2014-2023 period reveals clear spatial disparities. The hardest-hit areas form a central corridor extending through the city's north-south axis, where both outage frequency and duration are elevated, while the outer quadrants are comparatively less affected (Fig. 3a, b). PV-only system adoption is more concentrated in the urban core and in the eastern and southern suburbs (Fig. 3c). PV-plus-storage deployment is also uneven, with modest clustering observed in southern and eastern neighbourhoods (Fig. 3d). To better illustrate how reliability stress can be highly localized and transient, we examine a series of outage events on April 3, 2018, when severe thunderstorms swept across the Midwestern United States, bringing high winds and heavy rainfall[54,55]. As shown in Supplementary Figure S2, outages first appeared in the north and west with limited customer loss (<5%) but intensified over several hours as the storm core shifted southward and westward, pushing several substations above 15% of customers offline.

The rapid rotation of localized outage hotspots underscores the importance of mapping reliability at finer spatial resolutions, such as the substation level used in this study, to accurately quantify outage exposure (duration and frequency) and estimate its effect on adoption.

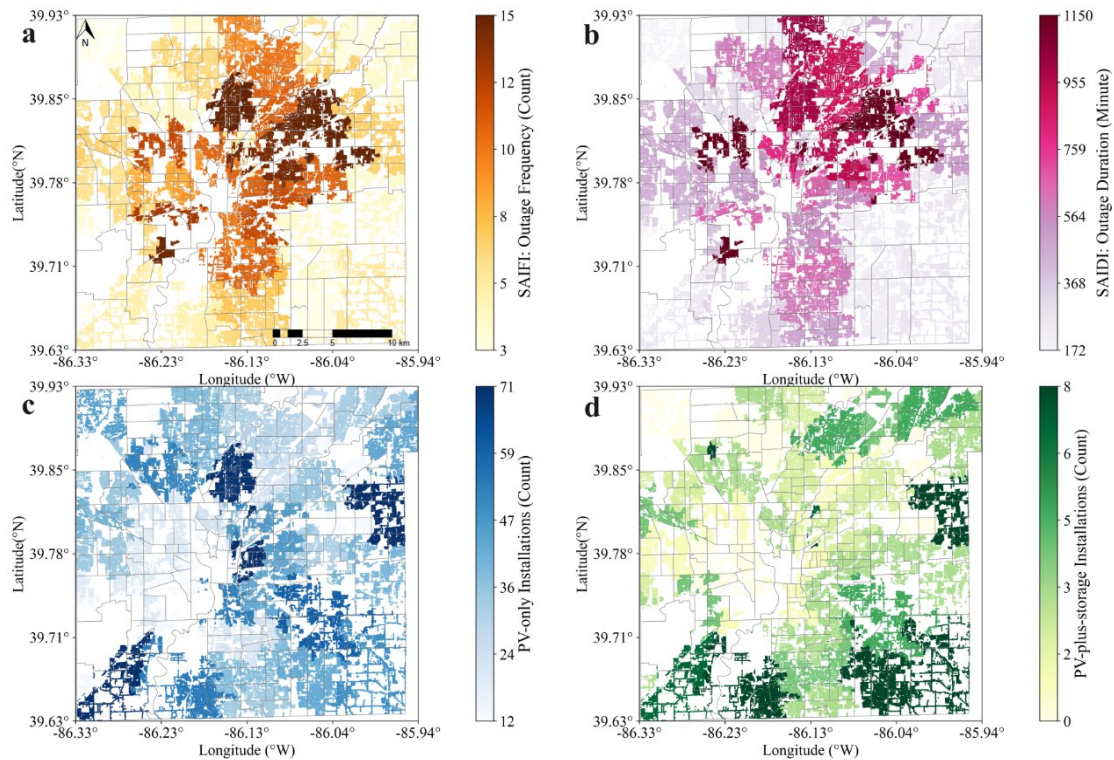

**Fig. 3 | Substation-level spatial patterns of aggregated power outages and DER installations. a,** Heatmap shows the distribution of cumulative frequency of power outages the average customer experienced, measured by SAIFI, between 2014 and 2023. **b,** Heatmap shows the distribution of cumulative duration of power outages the average customer experienced, measured by SAIDI, between 2014 and 2023. **c,** Heatmap shows the distribution of total number of installed PV-only systems by December 2023. **d,** Heatmap shows the distribution of total number of installed PV-plus-storage systems by December 2023. Each household is spatially aggregated using hexagonal binning, where each hexagon's colour intensity represents the sum of values for all households within its geographic area. Longitude and latitude coordinates are displayed in degrees. Census tract boundaries are shown as grey lines to provide spatial context. Blank areas on the map correspond to non-residential areas, which were excluded from the analysis (see Methods).

**Power outages reduce the adoption of PV-only systems**

We develop a two-part econometric panel model using monthly adoption data at the substation level. The outcome of interest is the quarterly new-installation rate of PV-only systems, defined as the share of newly installed PV-only systems in the preceding three months normalized by the total number of households served by each substation. The baseline models (see Table 1) incorporate all outage events to provide a comprehensive assessment of outage impact. To isolate the effects of longer power outages, we also construct alternative outage duration and frequency metrics by excluding events shorter than 1 hour.

The two-part model consists of a logistic regression (first part) and a generalized linear model (GLM) with a log link and Gaussian distribution (second part). The logistic regression estimates the probability that at least one new PV system is installed in a given substation-month, representing a transition from zero to positive adoption. The GLM estimates the magnitude of new installations, conditional on adoption occurring. Our model controls for various known predictors influencing residential solar PV adoption identified in the literature (see Supplementary Table 1), including socioeconomic characteristics, policy incentives, and peer effects. Summary statistics of key variables are reported in Supplementary Table 2. To account for unobservable confounders, we include both quarter and substation fixed effects, which control for time-varying and location-specific heterogeneity.

To address potential endogeneity, we use wind gust speed as an instrumental variable (IV) for outage duration (see Methods). This IV approach enables causal identification important because endogeneity may arise from reverse causality and omitted variable bias, which may not be fully addressed by the quarterly and substation fixed effects. One cause of possible endogeneity is that the adoption of solar PV systems can influence outage frequency or duration by increasing voltage and load variability during peak generation period[56,57]. Another cause is that unobserved factors, such as aging distribution infrastructure or variations in tree trimming, may simultaneously affect outage rates and solar PV adoption[58,59]. We find that wind gust speed is a strong and relevant instrument (F-statistic = 14.65), as nearly half of power interruptions in Indianapolis are related to high-wind events. Gust speed is also exogenous to PV adoption, in the sense that wind conditions are unlikely to influence household decisions to install solar PV. While extreme winds can theoretically damage solar panels, such events are rare in Indianapolis. Thus, wind speed is a valid and strong instrument for power outage duration (SAIDI). For outages lasting more than one hour, wind-gust speed remains a strong instrument for outage duration (F-statistic = 14.08) and is marginally acceptable (F-statistic = 10.08, slightly above the conventional threshold of 10) for outage frequency (SAIFI). But when we include all outage events, wind-gust speed is no longer strong enough for outage frequency, with an F-statistic of only 8.22. One possible explanation is that while strong winds are a primary cause of long-duration outages, the number of discrete events is also influenced by other factors, such as wildlife interference, vegetation contact, or operational switching, that are unrelated to wind conditions. The IV regression results are shown in Supplementary Note 1.

Table 1 reports the regression results from the two-part model when considering all events. In the first part (logistic regression), the coefficient on 3-month outage duration (SAIDI) is positive and statistically significant (0.079, $p < 0.01$), indicating that long and recent outages increase the probability of any PV adoption occurring within a substation three months after an outage. Quantitatively, each additional minute of outage duration in the prior quarter increases the odds of any adoption rate (relative to non-adoption) by 8.2%. When the analysis is restricted to major outages lasting more than one hour, the estimated coefficients become larger (see Supplementary Table 3), consistent with the hypothesis that more severe and salient disruptions prompt adoption responses. By contrast, coefficients on outage frequency (SAIFI) are negative for the 3- and 6-month horizons, suggesting that frequent short outages reduce the probability of adoption. This pattern implies that while prolonged outages can trigger adoption interest, recurrent minor disruptions have a deterrent effect. However, as outage frequency is not instrumented in these specifications, this effect should be interpreted with caution.

In the second part (using the Generalized Linear Model, or GLM), coefficients of outage duration are consistently negative and significant across all specifications, indicating that, conditional on some adoption occurring, longer outage durations are associated with lower new-installation rates within the substation. This divergence suggests that prolonged outages act as a trigger for adoption in non-adopting areas but deter further uptake in adopting areas. Moreover, the magnitude of the negative effect of outage duration decreases as the measurement horizon lengthens, implying that exposure to longer outage events exerts a weaker influence on adoption intensity than short-term disruptions. Most coefficients for outage frequency remain statistically insignificant across models.

| | Quarterly PV-only system new-installation rate | | | |
|---|---|---|---|---|
| | (1) | (2) | (3) | (4) |
| **Logistic regression (Part 1)** | | | | |
| 3-month SAIDI | 0.079** | | | |
| | (0.027) | | | |
| 6-month SAIDI | | 0.008 | | |
| | | (0.001) | | |
| 9-month SAIDI | | | -0.008 | |
| | | | (0.007) | |
| 12-month SAIDI | | | | -0.009 |
| | | | | (0.007) |
| 3-month SAIFI | -0.642* | | | |
| | (0.305) | | | |
| 6-month SAIFI | | -0.413* | | |
| | | (0.207) | | |
| 9-month SAIFI | | | -0.086 | |
| | | | (0.149) | |
| 12-month SAIFI | | | | -0.071 |
| | | | | (0.115) |
| Constant | -1.890 | -4.939 | -6.172 | -6.758 |
| | (7.665) | (7.501) | (7.293) | (6.955) |
| Controls | YES | YES | YES | YES |
| Quarter*Substation FE | YES | YES | YES | YES |
| Observations | 4,884 | 4,752 | 4,664 | 4,576 |
| Pseudo R2 | 0.535 | 0.538 | 0.536 | 0.539 |
| **GLM (Part 2)** | | | | |
| 3-month SAIDI | -0.026*** | | | |
| | (0.003) | | | |
| 6-month SAIDI | | -0.011*** | | |
| | | (0.001) | | |
| 9-month SAIDI | | | -0.008*** | |
| | | | (0.001) | |
| 12-month SAIDI | | | | -0.007*** |
| | | | | (0.001) |
| 3-month SAIFI | -0.003 | | | |
| | (0.029) | | | |
| 6-month SAIFI | | 0.007 | | |
| | | (0.023) | | |
| 9-month SAIFI | | | -0.021 | |
| | | | (0.024) | |
| 12-month SAIFI | | | | -0.027* |
| | | | | (0.011) |
| Constant | -7.835*** | -7.638*** | -7.625*** | -7.894*** |
| | (0.910) | (0.956) | (0.964) | (0.975) |
| Controls | YES | YES | YES | YES |
| Quarter*Substation FE | YES | YES | YES | YES |
| Observations | 2469 | 2437 | 2404 | 2372 |

**Table 1 | Regression results of the two-part model showing the impact of power outages on residential PV-only system new installation at the substation level.** This table reports the estimated coefficients and standard errors (in parentheses). Standard errors are clustered at the substation level. Significance levels: *$p < 0.05$, **$p < 0.01$, ***$p < 0.001$. All specifications include substation-quarter fixed effects and controls.

In the logistic regression (see Table 1), longer recent outages increase the likelihood of any adoption occurring. Major and recent outages may heighten awareness of reliability risks and prompt households to consider solar PV systems as a potential backup option. By contrast, in the GLM, which examines substations where at least one installation occurred, longer outage durations are negatively associated with the intensity of adoption, i.e., with lower PV adoption. This suggests that prolonged outages may deter additional uptake. Several possible mechanisms could explain this pattern: households may become more aware of the technical limitations of standalone PV systems during blackouts, or they may postpone investment in anticipation of grid upgrades or invest in backup solutions such as diesel generators[60].

Because the two components of a two-part model can exert opposing effects, average marginal effects (AMEs) are used to capture their combined causal influence on the adoption outcome. Figure 4a presents the AMEs from the two-part model, estimating the effect of a one-unit increase in outage duration or frequency on residential solar adoption. As the negative effect in the GLM outweighs the positive effect in the logistic part, the overall AME is negative, indicating that longer outages have been ultimately responsible for reductions in total installation rates across substations compared to the counterfactual scenario. We estimate that a one-hour increase in annual outage duration per household lowers the annual new-installation rate by approximately 0.012 percentage points. Given the low average baseline new-installation rate of 0.039% (2014-2023) in Indianapolis, this represents a substantial 31% decline relative to the historical mean between 2014 and 2023. By contrast, the effects of outage frequency are weaker. The AMEs for 3- and 12-month outage frequency are only marginally significant at the 10% level, and those for 6- and 9-month periods are statistically insignificant. The 95% confidence intervals for outage frequency estimates include zero (Fig. 4a), suggesting that outage frequency is less consistently associated with PV adoption than outage duration. Supplementary Table 4 reports the full set of AME values, standard errors, confidence intervals, and significance levels.

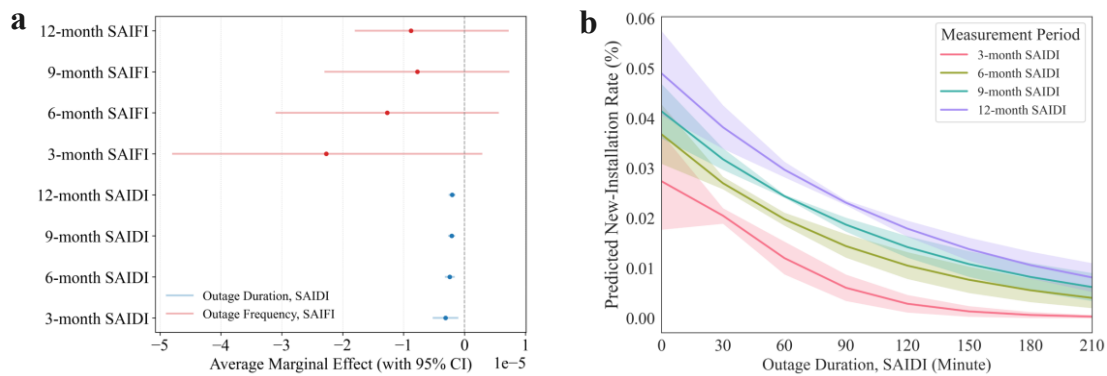

**Fig. 4 | Effects of outage duration and frequency on PV-only system adoption. a,** Marginal effects plot shows the combined average marginal effects of the two-part model for outage duration (SAIDI, all events) and outage frequency (SAIFI, all events). Points denote Average Marginal Effect estimates and bars represent 95% confidence intervals (CIs). The x-axis is scaled in scientific notation (1e-5). Negative AMEs indicate that longer outages are associated with lower subsequent installation rates of PV-only systems. **b,** Predictive margins plot shows estimated combined marginal effects of power outage duration on solar PV adoption across 4 model specifications with 95% CIs. Lines show marginal predictions from the two-part model of the unconditional quarterly new-installation rate by substation as SAIDI increases. Shaded bands represent 95% CIs. Curves differ by the SAIDI measurement horizon (1, 2, 3, or 4 quarters).

Figure 4b plots the calculated unconditional quarterly new-installation rate as a function of cumulative outage duration. Across all model specifications, the relationship is downward sloping, indicating that longer outage durations are consistently associated with lower installation rates after controlling for other factors. The steepest decline is observed for the 3-month outage duration, where predicted rates drop sharply as outage duration increases from 0 to about 90 minutes, after which the curve flattens. Longer measurement horizons display progressively flatter slopes, with the 12-month outage duration showing the smallest change in predicted adoption over the same range. This pattern is consistent with the AME results, suggesting that recent reliability disruptions exert a stronger negative influence on adoption decisions than outages distributed over longer periods. Although the shaded 95% confidence intervals widen at higher outage duration values, the negative association remains robust across all horizons.

In addition to power outages, several other factors also shaped PV adoption in the 12-month horizon model. Higher cooling degree days modestly increased the new-installation rate, reflecting stronger incentives to offset elevated cooling-related electricity demand. Economic drivers also play an important role. Higher retail electricity prices strengthen the financial case for solar PV adoption. Lower PV system costs are also strongly linked with a higher installation rate. In addition, a number of supportive policies play a significant role in shaping adoption outcomes. Net metering generates the strongest positive marginal effect, raising the annual new-installation rate by roughly 0.018 percentage points, equivalent to a 46% increase relative to the 2014-2023 baseline mean. The implementation of a PV installation easement law adds an additional increase of about 0.01 percentage points. The Residential Clean Energy Credit provides a smaller boost, increasing the annual new-installation rate by about 0.006 percentage points. Further discussion of these policy drivers is presented in the Discussion.

**Localized forecasting of future grid reliability and solar PV adoption under climate change**

Climate change increases the frequency and severity of weather-related hazards, including extreme heat, heavy precipitation, and high winds, that are among the leading causes of power outages[3]. We assess how climate change will shape future solar PV adoption by integrating climate-driven outage forecasts with the two-part model to produce neighbourhood-level projections under the RCP 4.5 scenario. This approach links high-resolution climate downscaling and event-level outage forecasts to household adoption behaviour, allowing us to identify how localized reliability changes shape adoption patterns across the city. Forecasts of this kind have not been achieved at comparable spatial and event-level resolution in earlier studies. The prediction uses a combination of historically observed daily meteorological data and Community Earth System Model projections downscaled to 7-kilometer resolution[47,61,62]. We are among the first to apply the hierarchical probabilistic conformal prediction (HPCP) to probabilistically forecast future outage occurrences at this level of spatial resolution (see Methods)[46]. We model outage occurrences using a point process framework[63]. The conditional intensity function is specified in a separable form: one component governs the spatiotemporal occurrences of outage events, which we model as a self-exciting Hawkes process; the other component governs the probability distribution of outage attributes, such as duration and number of customers affected, which we model using machine learning architectures that include softmax layers and feed-forward neural networks. Model parameters are estimated via maximum likelihood on a training dataset derived from historical outage records. The Hawkes process captures the clustering dynamics of outages and produces accurate and coherent predictions of future outages at an event-based level[64]. For more details on the downscaling of outages, see Methods section.

Figure 5 presents both temporal and spatial projections of outage duration and frequency under the RCP 4.5 scenario for global greenhouse gas concentrations from 2024 to 2040. To provide relatively conservative estimates, we focus on the RCP 4.5 emissions stabilization pathway. This intermediate scenario assumes that global mitigation efforts achieve some stabilization of greenhouse gas emissions, avoiding the most extreme warming outcomes while still producing significant climate impact. The use of the 50% prediction interval (25%-75%) narrows attention to the central range of model outcomes, excluding more extreme but less probable outage risks. Our estimates capture gradual climate-driven increases in outage duration but do not account for the sharp reliability spikes that can be associated with major events, which are evident in the historical record. More generally, our method for translating IPCC emissions scenarios into local storm and wind projections and quantifying their outage impacts can be used to estimate outage-related adoption effects under RCP 4.5 as well as under higher- and lower-temperature scenarios within the broader distribution of climate outcomes.

      Under this intermediate climate scenario, the model projects an upward trend in both outage duration and frequency (Fig. 5a, b). The probabilistic localized projection indicates that by 2040, the yearly average outage duration and frequency will reach around 495 minutes and 6.3 times, respectively, more than double the average annual outage duration and frequency between 2014-2023. Our downscaled model also suggests that power outages will become more widespread. In 2023, outages were concentrated in central and northern parts of the city (Fig. 5c). In 2040, the areas of elevated risk are projected to become far more widespread, extending across nearly all parts of the service territory. Some neighbourhoods will face both longer and more frequent outages, as reflected by the darker purple areas on the map (Fig. 5d). We then combine the projected outage trajectory for 2024-2040 with the estimated coefficients from the two-part model to assess the long-term impact of worsening grid reliability on future PV-only system adoption. To quantify this effect, we use the 2014-2023 average outage duration of 202 minutes as a counterfactual baseline in which reliability remains constant over time and assume a baseline annual new-installation rate of 0.039%, also corresponding to the historical mean between 2014 and 2023. Holding other factors constant, the results reveal a significant decline in new installations resulting from the lengthening of outages. If outage duration were to remain at the historical average, approximately 2,350 additional PV-only systems would be installed, absent any other changes, including policy, between 2024 and 2040 in a city of 377,726 households. Under the RCP 4.5 scenario, cumulative new installations are projected to decline to around 650 systems, representing a 72% reduction relative to the counterfactual scenario.

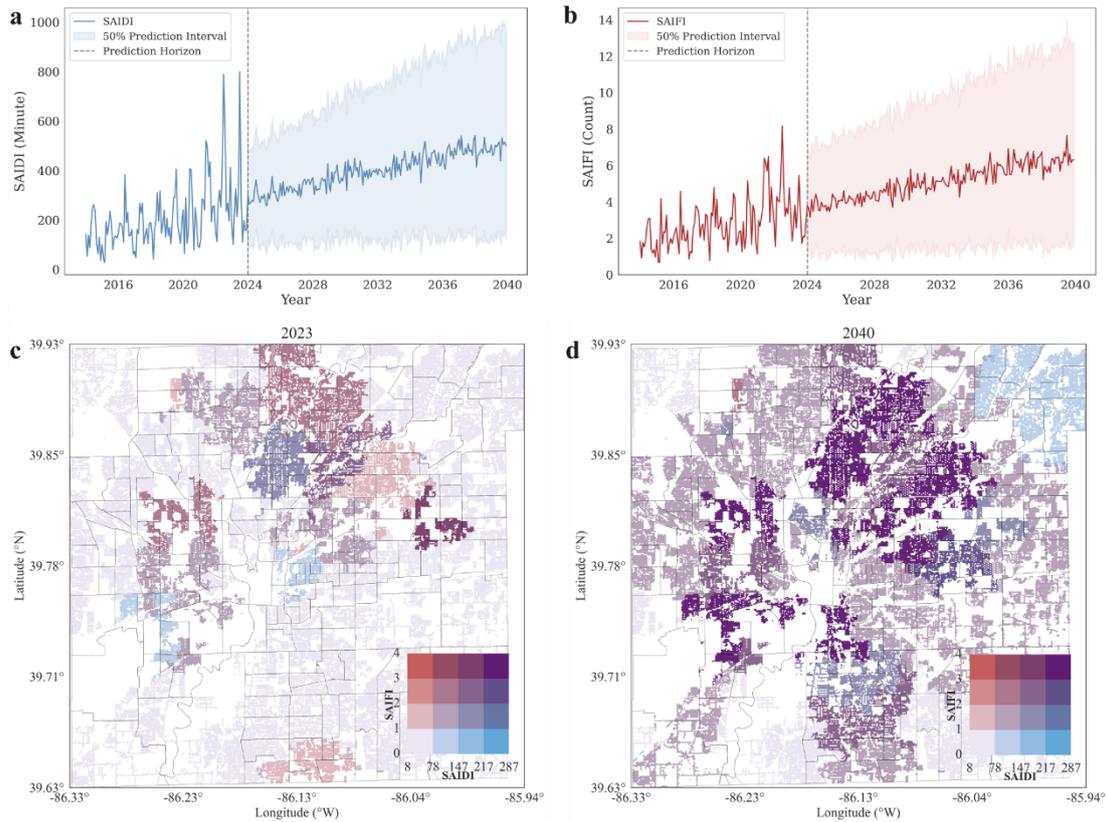

**Fig. 5 | Changes in outage duration and frequency under RCP 4.5 scenario. a-b,** Historical and projected trends of SAIDI and SAIFI under RCP 4.5, respectively. The solid lines show observed values (2014-2023) and predicted values (2024-2040). The dash lines separate the observed and predicted period. The shaded bands represent the 25%-75% prediction intervals. Results indicate a gradual increase in outage duration and frequency, with widening uncertainty bands over time. **c,** Bivariate hexbin map of yearly grid reliability indices in Indianapolis for 2023. Colours are assigned using a 4×4 bivariate colour scale, where increasing blue intensity corresponds to higher cumulative outage durations and increasing red intensity corresponds to higher outage frequencies. Purple colours indicate areas where both metrics are elevated. **d,** Bivariate hexbin map of yearly grid reliability indices in Indianapolis projected for 2040. Longitude and latitude coordinates are displayed in degrees. Each hexbin represents the cumulative value of the variable of interest, e.g., outage frequency, for substations falling within the corresponding cell. Census tract boundaries are shown as grey lines to provide spatial context. Blank area on the map correspond to non-residential area, which were excluded from analysis.

## 3    Discussion

Our findings show that longer power outages deter the adoption of grid-connected PV-only systems, a negative effect that has not been documented in prior studies. This suggests that prolonged power outages may trigger active information-seeking and belief-updating behaviour, during which households learn that residential solar systems without islanding capability or storage are designed to shut down during outages to prevent back-feeding and protect utility workers, and therefore cannot provide backup power[65]. During California's Public Safety Power Shutoffs, many homeowners and building operators were surprised to discover that their rooftop solar systems ceased operating once the shutoffs began[66,67]. Similar to positive peer effects[68], negative information about unattractive product features can also become salient through social interactions, deterring technology diffusion[69–72]. One recent UK study shows that when financial incentives for solar adoption decline, peer effects may turn negative, as critical discussions among neighbours reduce the perceived attractiveness of installation[72]. We provide evidence that negative peer effects may also emerge during prolonged outages, as households may draw negative inferences about PV-only systems through word-of-mouth interactions with early adopters.

Together, these mechanisms suggest that power outages do not uniformly encourage or discourage energy technology adoption. Their effects depend on how technologies perform under reliability stress. Whereas PV-only systems offer no backup capability, other distributed technologies that enhance energy self-sufficiency, such as battery storage and generators, can become more attractive in the same context. One paper finds that a one standard deviation increase in outage hours boosts monthly battery storage capacity by 32% within 3-5 months in California[33]. Similarly, an increase of 10 hours in annual outage duration was found to raise the PV-plus-storage adoption shares by 22%[34]. Another study shows that loss aversion and defensive expenditures incentivized households to adopt backup generators in outage-prone regions across the US in anticipation of storms[60]. In contrast, technologies that depend on reliable grid access, such as electric vehicles (EVs), experience reduced adoption under frequent outages, as charging disruptions erode consumer confidence and heighten range anxiety. One study in China found that each additional power outage in a city within a month is associated with roughly a 1% decline in new EV registrations while boosting sales of internal combustion engine vehicles.

Our findings contribute to this literature by providing highly granular evidence and a credible causal identification strategy in the case of residential PV systems and reveal a potential vicious cycle: as climate change intensifies extreme weather events and increases the risk of power outages, households become less likely to adopt PV-only systems, slowing decarbonization and reinforcing exposure to climate risks. Breaking this cycle requires policies that simultaneously enhance energy-system resilience, by further facilitating combined installation of solar and storage or by additional incentives for solar, for instance, and promote low-carbon transitions. Among all policy variables in our model, net metering exerted the strongest positive marginal effect on PV-only adoption, highlighting its central role in household investment decisions[73]. The implementation of net metering, effective from 2005 to 2022[74], raised the annual new-installation rate by about 0.02% over the 2014-2023 period. The Indiana Solar Easement Law (House Enrolled Act 1196)[75], which prohibits homeowners' associations from restricting rooftop solar installations since 2022 in most circumstances, further increased the rate by 0.01%, emphasizing the importance of removing regulatory and siting barriers to deployment.

However, the policy support for residential DERs is weakening. At the federal level, the Residential Clean Energy Credit will be phased out by the end of 2025[76], reducing the overall financial attractiveness of residential DERs. At the state level, Indiana repealed the mandate for utilities to credit solar generation at the retail rate and transitioned to net billing tariffs. All new systems installed after July 1, 2022 receive credits at roughly wholesale energy rates[74]. This transition in Indiana is part of a wider shift in residential solar incentives across the US and other developed economies, where net metering is being phased out in favour of net billing. More than a dozen US states, including Minnesota, Michigan, Indiana, Nevada, Connecticut, Kentucky, South Carolina, Iowa, and Illinois, have enacted legislation directing utilities to replace net metering with net billing[10]. Similar reforms have emerged across other developed economies: the UK introduced the Smart Export Guarantee to end retail-rate compensation in 2020[77]; Poland replaced net metering with wholesale-indexed net billing in 2022[78]; Cyprus requires all new systems to adopt net billing starting from 2026[79]; the Netherlands plans to phase out net metering entirely by 2027[16]. As many jurisdictions adopt similar tariff reforms, the economic attractiveness of grid-connected

PV-only systems is declining. The patterns we identify suggest that worsening outage conditions, when combined with decreasing remuneration for surplus solar generation, may further suppress uptake of PV-only systems in regions facing similar reliability challenges and evolving policy landscape. Our results also show that declining solar PV technology costs and rising electricity costs remain key drivers of technology diffusion[80,81]. A one-dollar-per-watt reduction in total installation costs increases the annual new-installation rate by 0.02%, equivalent to a 50% rise relative to the baseline mean. A one-dollar increase in retail electricity prices raises the rate by 0.006%.

These results highlight the need for policies that jointly strengthen grid resilience and sustain household incentives for DER adoption. Subsidies and tax incentives should prioritize retrofitting existing solar installations with batteries and encouraging the adoption of PV-plus-storage systems. Tariff structures may be redesigned to reflect the resilience value of these behind-the-meter resources. Performance-based mechanisms, such as dynamic tariffs, time-of-use pricing, or capacity credits for peak-shaving, can reward households for reducing demand when the grid is stressed or for maintaining supply during outages[82,83]. Targeted incentives could also be offered in high-outage regions[10], for example by maintaining net metering or offering location-specific tax rebates. Regulatory reforms should further remove non-financial barriers by streamlining interconnection procedures and standardizing permitting, which can reduce the transaction costs that deter adoption. Continued public and private investment in lowering the costs of solar and battery technologies will enhance affordability and expand access, particularly for lower-income households[84].

For electricity regulators and utilities in general, failing to account for the deterrent effects of outages risks systematic overestimation of future residential solar uptake and associated emissions reductions, leading to inefficient grid investment and planning decisions. Improving the reliability of distribution networks is also essential to prevent further disincentives for PV-only adoption. Utilities should use the approach in this paper to leverage granular outage and smart meter data to identify reliability hotspots, prioritize network hardening, and apply predictive maintenance[85]. Finally, enhancing consumer awareness through public education on system functionality, backup options, and retrofit technologies can align expectations and foster informed, resilience-oriented adoption decisions.

As noted, the lessons from Indianapolis extend beyond the city. Much of the Midwestern US, particularly within the Midcontinent Independent System Operator region, shares similar structural features: centralized investor-owned utilities, regulated retail markets, aging thermal generation fleets, and growing exposure to extreme weather events. Internationally, comparable challenges exist in other advanced economies where residential PV adoption is increasing while network hardening and resilience planning lag behind climate impact. Recognizing that grid reliability and distributed adoption interact dynamically, energy policy must increasingly integrate resilience metrics alongside cost-effectiveness and equity to ensure that clean energy transitions remain both sustainable and robust under a changing climate.

This study focuses exclusively on PV-only systems and does not include other residential energy technologies, such as battery storage or heat pumps. Although household-level battery adoption data are available, uptake is limited in Indianapolis (fewer than 200 recorded installations in the 10-year period), preventing robust statistical inference. The utility dataset captures only cases of simultaneous PV-plus-storage adoption, potentially omitting households that added batteries after their initial PV installations. This limitation constrains the ability to analyse sequential adoption dynamics, such as how prior PV ownership or exposure to power outages influences subsequent investment in storage. Future research should integrate longitudinal data that track retrofit and upgrade behaviour over time, enabling more comprehensive modelling of the residential energy system evolution. Expanding the analytical scope to include emerging technologies such as heat pumps, EVs, and smart energy management systems would further illuminate how multi-technology adoption pathways contribute to household resilience and decarbonization under increasing climate and grid stress.

## 4    Methods

**Power outage data and calculation of SAIDI and SAIFI**

This study uses proprietary data on power outage events provided by AES Indiana, the only investor-owned utility that serves the Indianapolis region. These data encompass all outages at the circuit level from January 2014 to December 2023, offering detailed insights into each event, including the location, start and end times, causes, and the number of customers affected. We calculate the SAIDI and SAIFI following the IEEE Standard 1366 guidelines for electric power distribution reliability indices. The calculation includes outages occurring on major event days, such as those caused by storms and other extreme weather events. We calculate the reliability metrics using the following two functions:

$$\text{SAIDI} = \frac{\sum_{i=1}^{N} U_i N_i}{N_T} \qquad (1)$$

$$\text{SAIFI} = \frac{\sum_{i=1}^{N} N_i}{N_T} \qquad (2)$$

$U_i$ is the duration of outage event $i$ (in minutes), calculated as the difference between recorded start and end times. $N_i$ is the number of customers affected by event $i$. $N_T$ is the total number of customers served within the substation at the start of the period. $N$ is the number of outage events observed within the period. The baseline model is based on the full sample of outage events. For additional specifications, we filter out outage events lasting longer than 1 hour. The SAIDI and SAIFI values for these specifications are computed based on these filtered events. Results were validated against reported indices to the IURC[86].

In addition to outage data, we incorporate geospatial information on the grid topology. This includes coordinates for individual customers (nodes), the configuration of distribution circuits, and the interconnections with substations. The topological data allow us to map outage events to affected areas with high accuracy, accounting for the hierarchical structure of the distribution network (node-circuit-substation). The outage and topological datasets provided by the local utility include all types of nodes in the distribution system, covering residential, commercial, and industrial customers. Given our study's focus on the relationship between household outage experiences and residential solar PV adoption, we restrict the analysis to residential service areas. To achieve this, we use the Indianapolis and Marion County Zoning Map Database and Indianapolis Zoning and Subdivision Ordinance (Chapter 742)[87,88], which provide the official classification and boundaries of land use within the city. We overlay these zoning boundaries of residential districts with the geospatial topology data to identify nodes that are located in residential districts. Supplementary Table 5 lists the dwelling districts included in the analysis.

**Technology adoption data**

This study uses proprietary interconnection records provided by the local utility, which are required for all grid-connected solar PV systems. Each installation must file an interconnection request before connection to the distribution network, making these records a comprehensive source of adoption data. The dataset is available at the household level, with geocoordinates of the installation and the exact date of interconnection at daily resolution. Each record also specifies the classification of the customer (industrial, commercial, or residential) as reported to the IURC, along with the circuit and substation to which the household is connected. We restrict the analysis to residential customers. While it is possible that a small number of fully islanded off-grid systems are not captured in the interconnection dataset, this number is expected to be negligible. Fully off-grid solar PV systems are generally not cost-effective in the Indianapolis context[89], since the absence of net-metering and grid interconnection would result in excess electricity being discarded rather than credited or sold back to the utility. We therefore treat the interconnection records as a near-complete record of residential PV adoption in the study area.

**Predictors of residential solar PV adoption**

We incorporate a set of predictors (covariates) of residential solar PV adoption identified in the literature. These include socioeconomic, demographic, dwelling, economic, and policy-related factors, as well as physical and infrastructure characteristics of the service territory.

Socioeconomic and demographic predictors are derived from the American Community Survey (ACS) at the census tract level[90]. These include income, educational attainment, average household size, gender distribution, age, and racial composition. We use utility customer density data from the local utility to weight and aggregate these tract-level statistics to substations. Dwelling characteristics, also sourced from the ACS at the census tract level, include the share of households using electricity for heating, median house value, dwelling type, and dwelling age bands. We measure peer effects as the installed base of PV systems within each substation, following the empirical approach proposed in the literature[68]. This variable captures social diffusion mechanisms whereby households are more likely to adopt if nearby peers have already done so. Economic covariates include total costs of residential solar PV systems and residential electricity prices. Cost data, encompassing both hard costs and soft costs, are obtained from the National Renewable Energy Laboratory[91]. Residential electricity prices are drawn from the U.S. Energy Information Administration's Form EIA-861M Monthly Electric Power Industry Report.

We identify three relevant policies using the Database of State Incentives for Renewables & Efficiency[92]. We create dichotomous indicators (0 for before implementation, 1 for after) based on effective dates. First, net energy metering, which allows customers to receive credits for excess solar generation exported to the grid at retail rates, was established in Indiana in 2004. The scheme underwent changes in 2017 via Senate Bill 309, phasing out retail-rate crediting by 2022 for new net billing tariffs[74]. Second, the Residential Clean Energy Credit, a federal tax credit under the Inflation Reduction Act, provides up to 30% of installation costs for systems placed in service after 2021, offsetting upfront expenses for homeowners[93]. Third, Indiana's Solar Easement & Rights Laws, updated in 2022 via House Bill 1196, prevent homeowners' associations from prohibiting solar panel installations and allow voluntary easements to protect solar access from neighbouring obstructions[75].

Environmental and infrastructure controls include weather variables such as average temperature, heating degree days (HDD) and cooling degree days (CDD), sourced from NOAA Applied Climate Information System[94]. These data capture climate-driven energy demand, as higher HDD or CDD may incentivize solar PV adoption for cost savings. Vegetation exposure is proxied by tree canopy cover from the U.S. Department of Agriculture Forest Service's Tree Canopy Cover (TCC) dataset[95]. We calculate annual substation-level average TCC using zonal statistics, taking the mean of raster pixels intersecting each substation polygon. Utility maintenance efforts are proxied by operation and maintenance expenditure reported by IPALCO Enterprises, the holding company of AES Indiana, in its quarterly financial statements (Form 10-Q) filed with the U.S. Securities and Exchange Commission[96]. Additionally, the number of utility poles per substation service area, provided by AES Indiana, serves as a proxy for overhead distribution exposure, with a higher pole count indicating less undergrounding and greater susceptibility to weather-related outages.

To address potential multicollinearity, we compute a correlation matrix (Fig. S3) and conduct a variance inflation factor test (Supplementary Note 2). For instance, high correlation between education and income led us to retain only income in baseline models. Supplementary Table 1 provides detailed variable descriptions and sources.

**IV construction from HRRR weather data**
Wind gust speed is a particularly strong predictor of power outage occurrence and severity. The mechanical force of wind rises with the cube of gust speed, and empirical studies show that the probability of powerline failure can increase sharply, with functional forms approaching a tenth-power relationship[51]. Evidence from major storms in U.S. and Europe suggests that wind gust speed is one of the most important predictors of power outage occurrence and severity[51,97]. To instrument for outage duration (SAIDI), we use wind gust speed from the High-Resolution Rapid Refresh (HRRR) model. The model provides hourly gridded weather fields at 3-kilometer resolution across the continental U.S. For each day in the study period (2014-2023), we retrieve the surface gust field at an hourly frequency using the Herbie Python package[45,98]. At each hour, we extract gridded gust speeds within the Indianapolis boundaries and overlay HRRR gust points with substation service area. For each substation $i$ on each day $d$, we identify the maximum hourly gust speed recorded within its polygon. To align with the quarterly frequency of reliability metrics, we then aggregate gusts to monthly and quarterly scales. We also construct 6-, 9-, and 12-month rolling sums to align with the cumulative SAIDI horizons used in the main specification. For quarter $t$ and horizon $h$ (e.g., 3, 6, 9, or 12 months), we sum the daily maximum gusts across all days $d$ within the $h$-month window preceding quarter $t$:

$$G_{i,t}^{(h)} = \sum_{d \in H(t,h)} max(gust_{i,d}) \qquad (3)$$

We conduct a comprehensive set of instrumental variable diagnostics to assess the strength and validity of the gust-speed instrument. These include cluster-robust first-stage F-tests, Kleibergen–Paap statistics, partial $R^2$, and weak-IV-robust Anderson-Rubin tests. The results, reported in Supplementary Note 1, confirm that the instrument is both relevant and valid, with first-stage F-statistics exceeding conventional thresholds and weak-IV-robust confidence sets that exclude zero.

**Two-part model**
Two-part models have been widely used in various areas, such as medical expenditure, survival, and fertility analysis, for modelling a non-negative variable with a mass of zeros and a continuous distribution for values above zero[99,100]. Compared to areas of high solar PV penetration, such as California, many substations in Indianapolis recorded zero adoption within a given quarter. Fig. S4 illustrates the distribution of the quarterly solar PV adoption rate, characterized by a substantial proportion of zeros alongside a right-skewed continuous distribution of positive values. These zeros are substantive, reflecting genuine non-adoption rather than censoring, and stem from multiple constraints. Notably, limited installer networks and low household awareness restricted adoption early in the study period. Additionally, the COVID-19 pandemic disrupted installations through supply-chain bottlenecks, permitting delays, and labour shortages. Traditional transformations of the dependent variable, such as log-plus-constant and inverse hyperbolic sine, may introduce an arbitrary scale parameter, making estimates highly sensitive to that choice[101]. Simulation evidence shows that two-part models recover correct marginal effects on the natural scale without relying on such ad-hoc parameters, making them preferable to transformation-based approaches[101].

To estimate the causal effect of power outages on residential solar PV adoption, we use a two-part model consisting of a logistic regression and a GLM with a log link and gaussian distribution. The outcome is a 3-month adoption rate of residential PV, defined as the ratio of households adopting PV in the preceding three months to the total number of customers in substation $i$ at month $t$. Instead of monthly adoption rate, this empirical specification is justified by the fact that households require several months to evaluate options, find a contractor, apply for interconnection with utility, and install the system[102]. Supporting evidence comes from the California Solar Initiative program and Self-Generation Incentive Program, where the median time from request to installation is approximately 90 to 100 days[34,68]. The treatment variables are power outage duration and frequency, measured by SAIDI and SAIFI, respectively. These two metrics are constructed daily at the substation level and aggregated over rolling $h \in 3,6,9,12$ month periods. Consistent with research on the effects of exposure to extreme weather events[103,104], we hypothesize that more recent exposure to outages has a greater influence on PV adoption. In addition, by modelling duration and frequency separately, the study compares the effects of prolonged but rare outages with frequent yet brief disruptions.

The first stage regression predicting outage duration $S_{i,t}^{(h)}$ using wind gust speed. The predicted values are then used in both parts of the model. The IV function is as follows:

$$S_{i,t}^{(h)} = \alpha_0 + \alpha_1 G_{i,t}^{(h)} + \alpha_2 X_{i,t} + \lambda_t + \mu_i + \epsilon_{i,t} \quad (4)$$

$S_{i,t}^{(h)}$ is cumulative SAIDI over horizon $h$, $X_{i,t}$ are controls, $\lambda_t$ is time fixed effect, $\mu_i$ is substation fixed effect, and $\epsilon_{i,t}$ is the residual. Standard errors are clustered at the substation level.

In the second stage, we use logistic regression models to estimate the likelihood that a substation adopts at least one residential solar PV system in a given quarter (first part). $Y_{i,t}$ denotes 3-month solar PV adoption rate. $D_{i,t} = 1 Y_{i,t} > 0$ indicates whether substation $i$ in time $t$ records any adoption in the preceding three month. $F_{i,t}^{(h)}$ denotes cumulative SAIFI over horizon $h$. Standard errors are clustered at the substation level. The first part logistic regression is as follows:

$$ln \frac{\Pr(D_{i,t}=1)}{1-\Pr(D_{i,t}=1)} = \beta_0 + \beta_1 S_{i,t}^{(h)} + \beta_2 F_{i,t}^{(h)} + \beta_3 X_{i,t} + \beta_4 \widehat{\epsilon_{i,t}} + \lambda_t + \mu_i \qquad (5)$$

Conditional on $D_{i,t} = 1$, the second-part uses a GLM with a log link function and gaussian distribution to model the conditional mean. We cluster standard errors by substation and use the same fixed effects and covariates as in the first part. The second part GLM regression is as follows:

$$ln\left(\mathbb{E}[Y_{i,t}|D_{i,t}=1]\right) = \gamma_0 + \gamma_1 S_{i,t}^{(h)} + \gamma_2 F_{i,t}^{(h)} + \gamma_3 X_{i,t} + \gamma_4 \widehat{\epsilon_{i,t}} + \lambda_t + \mu_i \quad (6)$$

**CESM data and hierarchical probabilistic conformal prediction model**
Our prediction work relies on a combination of historically observed daily meteorological data (1915–2013) across Indiana and statistically downscaled climate model projections. Specifically, we use the Community Earth System Model-Community Atmosphere Model (CESM1-CAM5) outputs under the RCP 4.5 scenario (2011–2040), downscaled to 1/16° resolution at a daily time step. The dataset includes precipitation, near-surface air temperature, daily maximum and minimum temperature, and wind speed. These variables are aggregated to the substation level and fed into the outage forecasting model, ensuring that both historical variability and projected climate trends are incorporated into the prediction of outage occurrences.

We apply conformal prediction[105,106], an uncertainty quantification method with strong finite-sample statistical guarantees, to estimate the prediction error. This involves generating multiple simulated trajectories from the point process using the thinning algorithm[63], and then computing a residual score for each observation. The residual score is defined as the minimum absolute difference between the observed outcome and the set of simulated predictions across repeated runs. For any future time step, we construct prediction intervals at each substation. These intervals are defined by taking the minimum and maximum of the simulated predictions and then adjusting them using an empirical quantile of the residual scores. The point prediction is given by a single simulated trajectory from the point process model. This event-based structure allows for coherent prediction of not only when and where outages occur, but also how severe they are likely to be.

The prediction algorithm proceeds in three steps: First, we model the outage occurrences via a point process model, whose conditional intensity function is parameterized in the following separable form:

$$\lambda(t,s,m|\mathcal{H}_t) = \lambda g(t,s\,|\,\mathcal{H}_t) \cdot f(m\,|\,t,s,\mathcal{H}_t) \qquad (7)$$

$\lambda(t,s,m|\mathcal{H}_t)$ is defined as the conditional intensity of the process governing the time and location of outage occurrences, which we parameterize as a self-exciting Hawkes process with an exponential decaying kernel. The mark function $f(m|t,s,\mathcal{H}_t)$ models the conditional probability of outage duration and customer influence, which we parameterize using machine learning architectures consisting of softmax activation functions and recurrent neural network architectures. The model is fitted via maximum likelihood using a training dataset split from the observed dataset. Second, we carry out uncertainty calibration, which aims to estimate the expected error that the model makes across its predictions. This is done via drawing predictions from the point process model via a simulation procedure referred to as the thinning algorithm, and then collecting the minimum residual, i.e., non-conformity score, over $\mathcal{K}$ independent

simulations, defined as:

$$r_{i,s} = \min_{k=1,\ldots,K} \left| \hat{y}_{i,s}^{(k)} - y_{i,s} \right|, \ i = 1, \ldots, n \quad (8)$$

$n$ is defined as the number of total observed timesteps, $y_{i,s}^{(k)}$ is defined as the $k$-th prediction at the $i$-th timestep at substation $s$, and $y_{i,s}$ is its ground-truth counterpart. Third, for any future timestep $i' > n$, we define the prediction interval for substation $s$ as:

$$\text{PI}_{i',s} := \left[ \min_{k=1,\ldots,K} \hat{y}_{i',s}^{(k)} - \hat{Q}_s(1-\alpha), \max_{k=1,\ldots,K} \hat{y}_{i',s}^{(k)} + \hat{Q}_s(1-\alpha) \right] \quad (9)$$

$\hat{Q}_s$ is the empirical quantile function with $\{r_{i,s}\}_{i=1}^n$. The (mean) prediction is defined as a single simulated outcome from the point process model. We include the detailed and more rigorous description of the complete procedure in Supplementary Note 3.

There are two main advantages to the event-based approach, i.e., self-exciting Hawkes process, that we adopt. First, the Hawkes process models inter-event dynamics through its excitation term, enabling us to capture outage clustering and cascading failures observed in the dataset, thereby improving the accuracy of outage modelling and prediction at a spatio-temporal level. On the other hand, regression models cannot explicitly capture these inter-event spatio-temporal dynamics, as they directly operate at the aggregated level. Second, event-based modelling approach facilitates the calculation of the downstream outage resilience index, as the model can encode outage characteristics such as duration and number of customers interrupted as "marks", which are then jointly modelled with the spatio-temporal dynamic. These marks, when predicted, can be simply aggregated to create resilience indices according to their definition. On the other hand, using prediction tools such as regression requires fitting and predicting these resilience indices separately, harming the coherence of the modelling process. Therefore, event-based modelling is not only interpretable but is also tailored for accurate prediction of outage events. We will further verify these intuitions by showing a baseline comparison experiment in Supplementary Note 4.

## 5    Data availability

The aggregated and desensitized data will be uploaded to the University of Cambridge Apollo repository for open access upon publication.

## 6    Code availability

All codes used in this analysis will be uploaded to the University of Cambridge Apollo repository for open access upon publication.

# 8  Acknowledgements

We thank Indianapolis Power & Light Co. dba AES Indiana for providing the data and valuable insights that greatly contributed to shaping this research. Part of this research was performed while the authors were visiting the Institute for Mathematical and Statistical Innovation (IMSI) for a workshop on "The Architecture of Green Energy Systems" in 2024, which was co-organized by L.D.A. and supported by the National Science Foundation (Grant No. DMS-1929348). We also thank Dr D. Cale Reeves, Dr Sergey Kolesnikov, Prof David Reiner, and seminar participants at the Cambridge EPRG Energy and Environment Seminar for their helpful comments and suggestions. L.D.A. is supported by a Senior Fellowship award from the JM Keynes Fellowship Fund at the University of Cambridge and by UKRI under the UK government's Horizon Europe funding guarantee (Grant No. 10062835) for the PRISMA project, which has also received funding from the European Union's Horizon Europe program (Grant No. 101081604). J.Z. is supported by the Cambridge Commonwealth European and International Trust.

# 9  Competing interests

The authors declare no competing interests.

## 10      Supplementary information

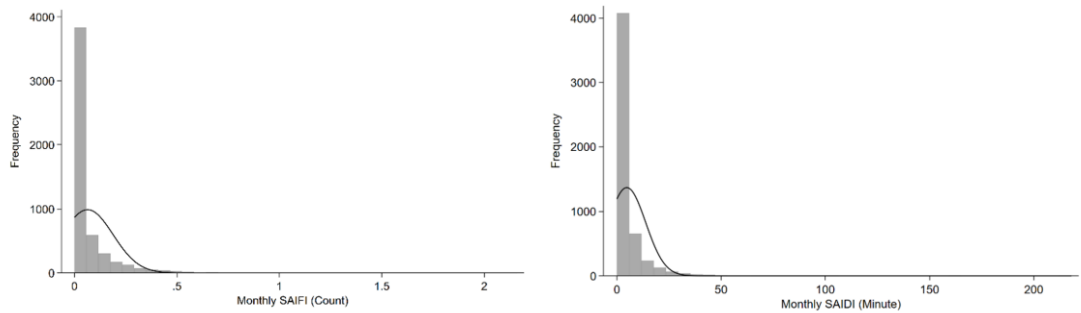

**Fig. S1 | Distribution of monthly outage indices across substations from 2014-2023.** Left panel: System Average Interruption Frequency Index (SAIFI, number of outages per customer). Right panel: System Average Interruption Duration Index (SAIDI, minutes of outage per customer). Both distributions are heavy-tailed: most substations record low outage levels in most months, while a small subset experiences severe disruptions.

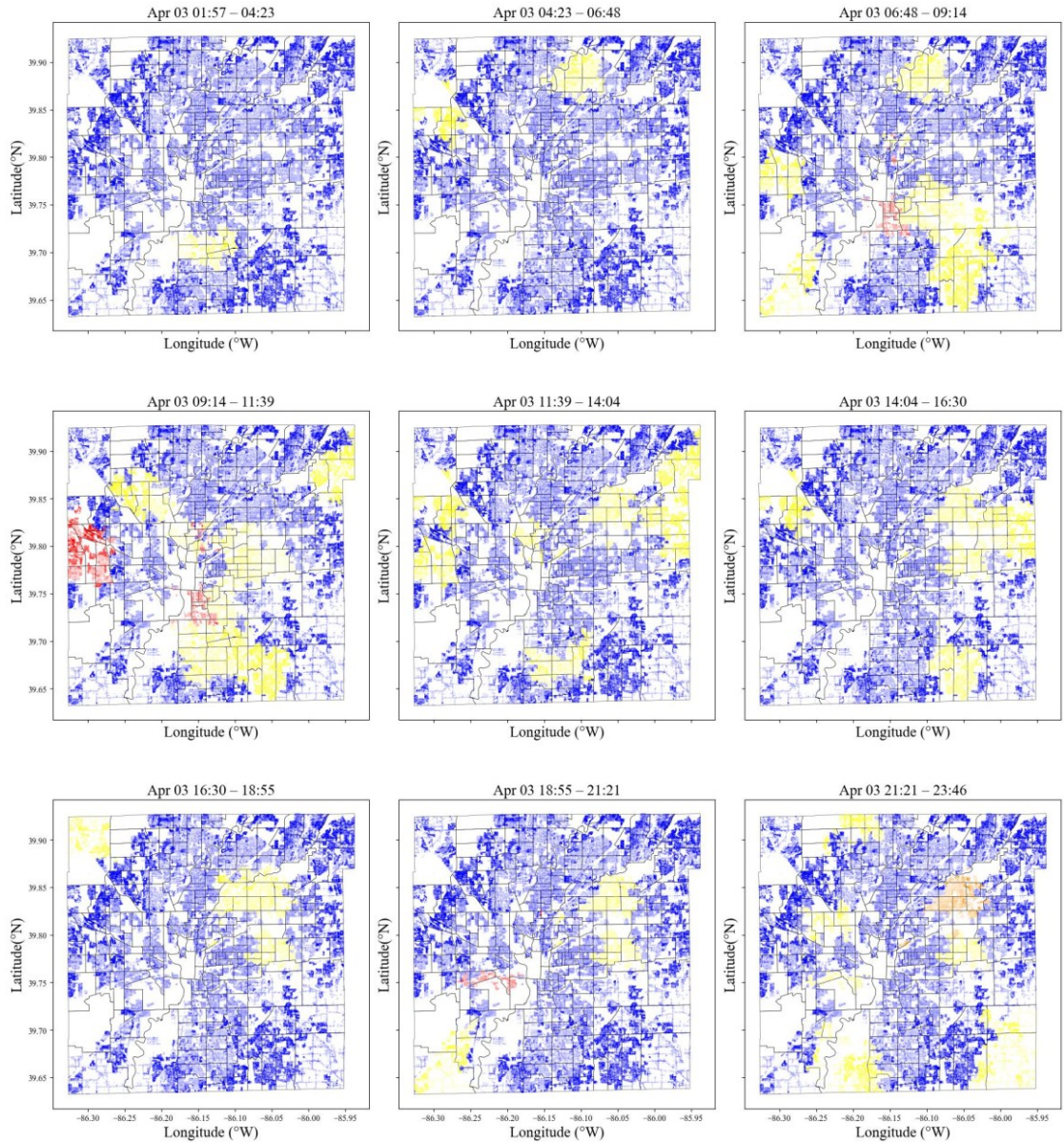

**Fig. S2 | Spatiotemporal evolution of power outages during the April 3, 2018 thunderstorm event in Indianapolis.** This 3×3 panel sequence illustrates the dynamic progression of power outages across Indianapolis substations during the severe thunderstorm event on April 3, 2018. Each panel represents a consecutive time interval throughout the day, with the temporal sequence reading left-to-right and top-to-bottom. Each dot represents a customer location. Outage severity is represented using a sequential colour scale: blue: 0% customer impact (no outages), yellow gradient: 0-5% customer impact (minor disruptions), orange gradient: 5-10% customer impact (moderate disruptions), red gradient: >10% customer impact (severe disruptions). Colour intensity within each range indicates proportional severity. Census tract boundaries33 are shown as grey lines to provide spatial context. Blank areas on the map correspond to non-residential areas, which were excluded from analysis.

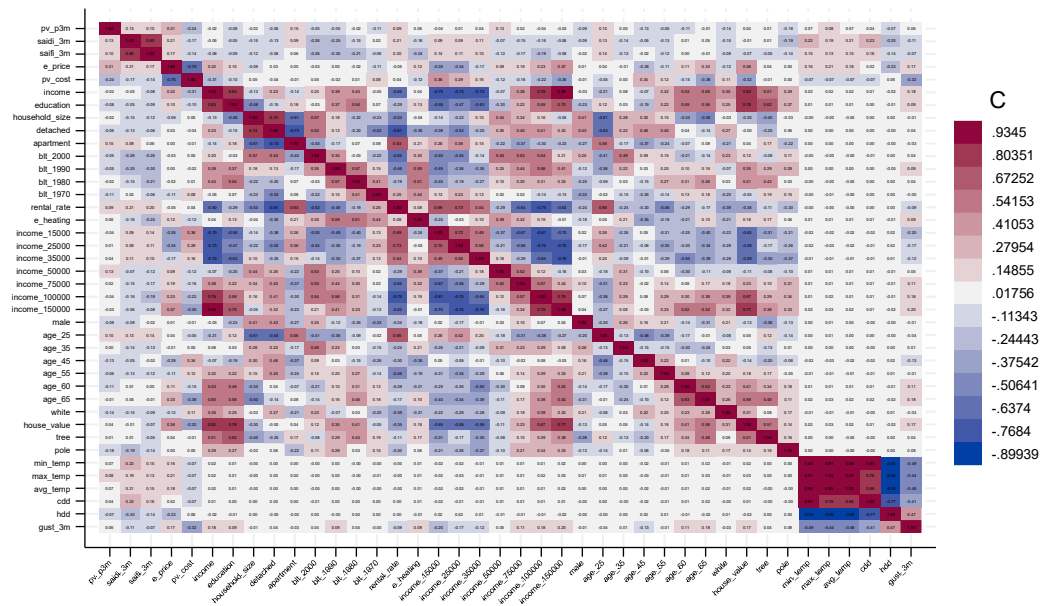

**Fig. S3 | Pairwise correlation heat map for all candidate continuous predictors.** The graph includes all candidate covariates. We examine the correlation among them using the Pearson correlation coefficient. Some of them are highly correlated (over the conventional threshold of ± 0.7), such as education level and median income, house price and rental rate. When two or more variables were highly correlated, we retained a single representative based on theory, data completeness, and parsimony. The final predictors are listed in Supplementary Note 2. We also verified multicollinearity using variance-inflation factors (VIF) on the estimation sample; all retained regressors met conventional VIF=10 thresholds. Binary and categorical variables (e.g., policy incentives before/after) are not shown here because Pearson correlations with dummies can be misleading and depend on marginal frequencies.

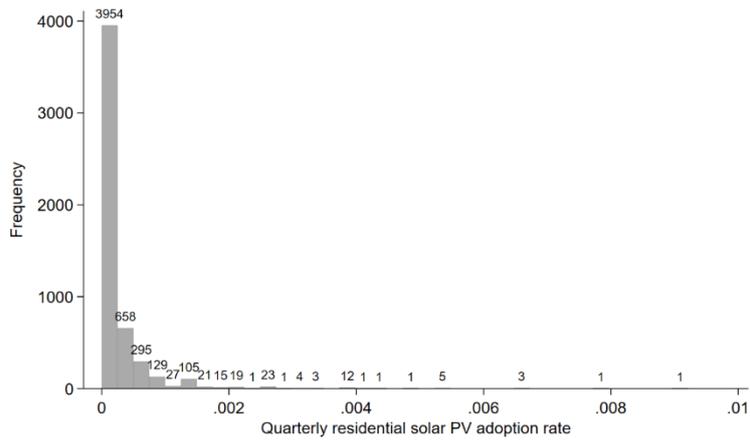

**Fig. S4 | Histogram of quarterly residential PV adoption rates across substation–quarters.** The distribution is zero-inflated: most observations are exactly zero, while the positive rates form a highly right-skewed tail. This excess-zero, overdispersed shape motivates a two-part approach, modelling the probability of any adoption separately from the conditional intensity given adoption.

| Category | Covariate | Frequency | Granularity | Source | Study |
|---|---|---|---|---|---|
| Socioeconomic and Demographics | Income: Median Household Income, Income Bands (%15,000 to $24,999, % $25,000 to $34,999, % $35,000 to $49,999, % $50,000 to $74,999, % $75,000 to $99,999, % $10,0000 to $14,9999, % $150,000 to $199,999) | Yearly | Census tract | American Community Survey | (Barbose et al., 2021; Bollinger & Gillingham, 2012; Shakeel et al., 2023; Wolske, 2020) |
| | Education: % bachelor's degree or higher (population 25 years and over) | Yearly | Census tract | American Community Survey | |
| | Median household size | Yearly | Census tract | American Community Survey | |
| | Gender: % of male | Yearly | Census tract | American Community Survey | |
| | Age bands: % 25 to 29, % 35 to 39, % 45 to 49, % 55 to 59, % 60 to 64, % 65 to 69) | Yearly | Census tract | American Community Survey | |
| | Race: % of white | Yearly | Census tract | American Community Survey | |
| Dwelling Characteristics | Electric heating: % Electricity as heating fuel | Yearly | Census tract | American Community Survey | (Bollinger & Gillingham, 2012; Dimitriou et al., 2024; Qiu et al., 2017; Wee, 2016) |
| | Renter-occupied rate | Yearly | Census tract | American Community Survey | |
| | Dwelling value: Median value | Yearly | Census tract | American Community Survey | |
| | Dwelling type: % detached, % apartment | Yearly | Census tract | American Community Survey | |
| | Dwelling age: years structure built (% built 2000 to 2009, % built 1990 to 1999, % built 1980 to 1989, % built 1970 to 1979) | Yearly | Census tract | American Community Survey | |
| Belief | Belief in climate change | Yearly (except for 2015 and 2017) | County | Climate Change in the American Mind National Survey | (Mildenberger et al., 2019; Schulte et al., 2022) |
| Peer Effects | Installed base: number of residential solar PV installed within each substation | Monthly | Substation | AES Indiana | (Bollinger & Gillingham, 2012; Curtius et al., 2018; Graziano et al., 2019; Rai et al., 2016; Reeves et al., 2017) |
| **Price** | Costs of solar PV system | Yearly | National | NREL Solar Technology Cost Analysis | (Chesser et al., 2018; Islam & Meade, 2013; Lemay et al., 2023) |
| | Residential electricity price | Monthly | State | EIA Form EIA-861M Monthly Electric Power Industry Report | |
| **Policies** | Net metering scheme (ended in June 2022) | | State | DSIRE Policy Database | (Lemay et al., 2023; O'Shaughnessy et al., 2021; Shakeel et al., 2023) |
| | Net billing tariffs (effective from July 2022) | | State | DSIRE Policy Database | |
| | Residential Clean Energy Credit | | State | DSIRE Policy Database | |
| | Indiana Solar Easements & Rights Law | | State | DSIRE Policy Database | |
| **Others** | Average temperature | Monthly | County | NOAA Applied Climate Information System | (Fikru & Gautier, 2015; Lamp, 2023; Shi et al., 2024) |
| | Heating and cooling degree days | Monthly | County | NOAA Applied Climate Information System | |
| | Tree canopy cover | Yearly | 30 meters | USDA Forest Service | |

**Supplementary Table 1 | Initial selection of covariates.** Variables are grouped by category and listed with their temporal frequency, spatial granularity, sources, and key references.

| Variable | Mean | SD | Min | Max |
|---|---|---|---|---|
| Monthly SAIDI | 4.75 | 9.03 | 0 | 218.01 |
| Monthly SAIFI | 0.06 | 0.12 | 0 | 2.15 |
| 3-month SAIDI | 14.17 | 18.13 | 0 | 244.31 |
| 3-month SAIFI | 0.19 | 0.24 | 0 | 3.09 |
| 6-month SAIDI | 28.04 | 29.38 | 0 | 340.46 |
| 6-month SAIFI | 0.37 | 0.38 | 0 | 3.73 |
| 9-month SAIDI | 41.17 | 39.16 | 0 | 435.70 |
| 9-month SAIFI | 0.56 | 0.52 | 0 | 4.48 |
| 12-month SAIDI | 53.87 | 48.62 | 0 | 492.62 |
| 12-month SAIFI | 0.73 | 0.64 | 0 | 5.01 |
| Number of monthly installations | 0.28 | 0.70 | 0 | 12 |
| Number of 3-month installations | 1.12 | 1.64 | 0 | 18 |
| 3-month new installation rate | 0.0002 | 0.0005 | 0 | 0.009 |
| Median income | 73973.32 | 26866.98 | 31905.55 | 189465 |
| % income $35,000 to $49,999 | 14.63 | 3.18 | 4.25 | 22.71 |
| % income $75,000 to $99,999 | 11.75 | 3.59 | 3.521 | 22.83 |
| % income $10,0000 to $14,9999 | 12.73 | 5.93 | 1.70 | 35.95 |
| Household size | 2.55 | 0.31 | 1.56 | 3.27 |
| % male | 48.18 | 2.15 | 36.53 | 54.01 |
| % white | 61.68 | 17.43 | 24.21 | 91.69 |
| % age 25-29 | 15.89 | 3.27 | 6.06 | 33.32 |
| % age 35-39 | 13.02 | 1.65 | 8.70 | 18.6 |
| % age 55-59 | 2.55 | 0.31 | 1.56 | 3.27 |
| % age 55-59 | 6.47 | 1.19 | 3.5 | 11.63 |
| % age 65-69 | 7.45 | 1.97 | 2.9 | 15.33 |
| % electrical heating | 38.03 | 11.39 | 12.88 | 78.17 |
| % renter occupied | 40.70 | 16.23 | 4.25 | 83.6 |
| % detached | 64.43 | 17.30 | 16.7 | 70.96 |
| Median house value | 153191.10 | 71955.74 | 53700 | 588300 |
| % people believe in climate change happening | 68.74 | 2.31 | 64 | 70.96 |
| Number of solar PV installed in the same substation | 8.495 | 12.57 | 0 | 77 |
| Average temperature | 54.28 | 17.08 | 20.1 | 78.4 |
| Cooling degree days | 105.95 | 134.60 | 0 | 422 |
| % tree canopy cover | 18.31 | 5.57 | 10.66 | 35.33 |

**Supplementary Table 2 | Summary of descriptive statistics of key variables.** The mean, standard deviation, minimum and maximum values of key variables are summarized in this table.

| | Quarterly PV-only system new-installation rate | | | |
|---|---|---|---|---|
| | (1) | (2) | (3) | (4) |
| **Logistic regression (Part 1)** | | | | |
| 3-month SAIDI (events 1+ hour) | 0.091** | | | |
| | (0.033) | | | |
| 6-month SAIDI (events 1+ hour) | | 0.018 | | |
| | | (0.012) | | |
| 9-month SAIDI (events 1+ hour) | | | -0.005 | |
| | | | (0.008) | |
| 12-month SAIDI (events 1+ hour) | | | | -0.007 |
| | | | | (0.008) |
| 3-month SAIFI (events 1+ hour) | -1.239* | | | |
| | (0.533) | | | |
| 6-month SAIFI (events 1+ hour) | | -0.833* | | |
| | | (0.419) | | |
| 9-month SAIFI (events 1+ hour) | | | -0.372 | |
| | | | (0.287) | |
| 12-month SAIFI (events 1+ hour) | | | | -0.224 |
| | | | | (0.238) |
| Constant | -2.127 | -4.451 | -5.834 | -6.320 |
| | (7.525) | (7.344) | (7.278) | (6.98) |
| Controls | YES | YES | YES | YES |
| Quarter*Substation FE | YES | YES | YES | YES |
| Observations | 4,884 | 4,752 | 4,664 | 4,576 |
| Pseudo R2 | 0.534 | 0.538 | 0.537 | 0.539 |
| **GLM (Part 2)** | | | | |
| 3-month SAIDI (events 1+ hour) | -0.022*** | | | |
| | (0.003) | | | |
| 6-month SAIDI (events 1+ hour) | | -0.012*** | | |
| | | (0.001) | | |
| 9-month SAIDI (events 1+ hour) | | | -0.008*** | |
| | | | (0.001) | |
| 12-month SAIDI (events 1+ hour) | | | | -0.008*** |
| | | | | (0.001) |
| 3-month SAIFI (events 1+ hour) | 0.056 | | | |
| | (0.069) | | | |
| 6-month SAIFI (events 1+ hour) | | 0.016 | | |
| | | (0.032) | | |
| 9-month SAIFI (events 1+ hour) | | | -0.050 | |
| | | | (0.033) | |
| 12-month SAIFI (events 1+ hour) | | | | -0.005 |
| | | | | (0.018) |
| Constant | -7.521*** | -7.506*** | -7.563*** | -7.843*** |
| | (0.932) | (0.957) | (0.993) | (1.045) |
| Controls | YES | YES | YES | YES |
| Quarter*Substation FE | YES | YES | YES | YES |
| Observations | 2469 | 2437 | 2404 | 2372 |

**Supplementary Table 3 | Regression results (events longer than 1 hour) of the two-part model showing the impact of power outages on residential PV-only system new installation.** This table reports the estimated coefficients and standard errors (in parentheses). Standard errors are clustered at the substation level. Significance levels: *$p < 0.05$, **$p < 0.01$, ***$p < 0.001$. All specifications include substation-quarter fixed effects and controls.

| Variable | AME | Std. Error | 95% CI Lower | 95% CI Upper |
|---|---|---|---|---|
| Outage Duration (SAIDI) | | | | |
| 3-month SAIDI | $-3.12 \times 10^{-6}$** | $1.07 \times 10^{-6}$ | $-5.21 \times 10^{-6}$ | $-1.03 \times 10^{-6}$ |
| 6-month SAIDI | $-2.41 \times 10^{-6}$*** | $4.14 \times 10^{-7}$ | $-3.22 \times 10^{-6}$ | $-1.60 \times 10^{-6}$ |
| 9-month SAIDI | $-2.11 \times 10^{-6}$*** | $2.93 \times 10^{-7}$ | $-2.68 \times 10^{-6}$ | $-1.54 \times 10^{-6}$ |
| 12-month SAIDI | $-2.04 \times 10^{-6}$*** | $2.91 \times 10^{-7}$ | $-2.61 \times 10^{-6}$ | $-1.47 \times 10^{-6}$ |
| Outage Frequency (SAIFI) | | | | |
| 3-month SAIFI | $-2.27 \times 10^{-5}$* | $1.31 \times 10^{-5}$ | $-4.8 \times 10^{-5}$ | $2.94 \times 10^{-6}$ |
| 6-month SAIFI | $-1.27 \times 10^{-5}$ | $9.36 \times 10^{-6}$ | $-3.1 \times 10^{-5}$ | $5.63 \times 10^{-6}$ |
| 9-month SAIFI | $-7.74 \times 10^{-6}$ | $7.71 \times 10^{-6}$ | $-2.3 \times 10^{-5}$ | $7.36 \times 10^{-6}$ |
| 12-month SAIFI | $-8.78 \times 10^{-6}$* | $4.85 \times 10^{-6}$ | $-1.8 \times 10^{-5}$ | $7.28 \times 10^{-7}$ |

**Supplementary Table 4 | Average marginal effects of combined two-part model.** Significance levels: *$p < 0.05$, **$p < 0.01$, ***$p < 0.001$.

| Code | District Name |
|---|---|
| D-A | Dwelling Agriculture District |
| D-S | Dwelling Suburban District |
| D-1 | Dwelling District One |
| D-2 | Dwelling District Two |
| D-3 | Dwelling District Three |
| D-4 | Dwelling District Four |
| D-5 | Dwelling District Five |
| D-5II | Dwelling District Five-Two |
| D-6 | Dwelling District Six |
| D-6II | Dwelling District Six-Two |
| D-7 | Dwelling District Seven |
| D-8 | Dwelling District Eight |
| D-9 | Dwelling District Nine |
| D-10 | Dwelling District Ten |
| D-11 | Dwelling District Eleven |
| D-P | Planned Unit Development District |

**Supplementary Table 5 | Residential zones included in the study.** These zones are listed as residential areas in the Indianapolis-Marion County Consolidated Zoning and Subdivision Ordinance (City of Indianapolis, 2025b). Polygon files representing the residential zoning boundaries in Indianapolis and Marion County, Indiana were obtained from the IndyGIS Database (City of Indianapolis, 2025a).

**Supplementary Note 1:** IV regression results and statistical test

|  | (1) 3-month SAIDI (all events) | (2) 3-month SAIFI (all events) | (3) 3-month SAIDI (events 1+ hour) | (4) 3-month SAIFI (events 1+ hour) |
|---|---|---|---|---|
| 3-month Gust speed | 3.51*** | 0.03* | 3.11*** | 0.02** |
|  | (0.91) | (0.01) | (0.83) | (0.01) |
| Constant | 97.33*** | 1.40 | 97.33 | 0.57 |
|  | (25.49) | (0.48) | (2.88) | (3.21) |
| Controls | YES | YES | YES | YES |
| Substation-Month FE | YES | YES | YES | YES |
| Observations | 4884 | 4884 | 4884 | 4884 |
| R-squared | 0.36 | 0.27 | 0.35 | 0.23 |
| IV F-statistics | 14.65*** | 8.22* | 14.08*** | 10.08** |

Standard errors are clustered to the substation level. Standard errors in parentheses. *$p < 0.10$, **$p < 0.05$, ***$p < 0.01$.

We first examined instrument relevance by estimating the first-stage regression of outage duration on the gust instrument, including substation- and quarter-fixed effects, with standard errors clustered at the substation level. Instrument F-statistics are 14.65, 8.22, 14.08, and 10.08 for 3-month SAIDI (all events), 3-month SAIFI (all events), 3-month SAIDI (events 1+ hour), 3-month SAIFI (events 1+ hour) respectively. Except for 3-month SAIFI (all events), the F-statistics exceed the conventional Staiger–Stock threshold of 10. The first-stage coefficient on gust speed is also positive and statistically significant, consistent with the physical mechanism by which stronger wind gusts increase the likelihood of tree damage and powerline faults. To ensure inference was robust even under possible weak-IV scenarios, we also implemented the Anderson–Rubin (AR) test. For the three-month horizon, the AR test rejects the null of no effect at the 5% level ($\chi^2(1)=6.34$, $p=0.012$), and the associated AR confidence set excludes zero. The robustness of the AR results provides strong support for the validity of our causal estimates.

**Supplementary Note 2:** Selection of covariates and variance-inflation factors test

| Variable | VIF | 1/VIF |
|---|---|---|
| SAIDI_3m | 2.7 | 0.370872 |
| SAIFI_3m | 1.34 | 0.746953 |
| % detached house | 6.69 | 0.149464 |
| % house using electric heating | 6.22 | 0.160872 |
| House value | 4.04 | 0.247236 |
| Electricity price | 3.26 | 0.306661 |
| Peer effect | 2.99 | 0.334423 |
| PV cost | 2.77 | 0.361096 |
| % built 1970-1979 | 2.79 | 0.35862 |
| % built 1980-1989 | 3.94 | 0.254096 |
| % built 1990-1999 | 4.52 | 0.221403 |
| % built 2000-2009 | 5.21 | 0.192009 |
| % income_35000-4999 | 2.67 | 0.375103 |
| % income_50000-74999 | 2.62 | 0.382365 |
| % income_75000-9999 | 2.84 | 0.352714 |
| % income_100000-149999 | 6.23 | 0.160399 |
| % age_25-29 | 3.66 | 0.273386 |
| % age_35-39 | 1.97 | 0.507168 |
| % age_45-49 | 2.32 | 0.431768 |
| % age_55-59 | 2.05 | 0.488516 |
| % age_65-69 | 2.94 | 0.339618 |
| % white | 2.09 | 0.47859 |
| % tree canopy coverage | 2.02 | 0.494559 |
| % male | 1.79 | 0.558884 |
| Cooling degree days | 1.35 | 0.740673 |
| Mean VIF: 3.24 | | |

We computed variance-inflation factors (VIF) for the baseline covariates. All VIFs are below 7 (mean=3.24), comfortably under common rules of thumb of 10. The largest VIFs occur for detached housing share (6.69), share of household income 100k-150k (6.23), and share of houses with electric heating (6.22), reflecting expected overlap among housing type, income/wealth, and heating technology. Tolerance values (1/VIF) remain ≥0.15 for these variables, indicating that each still contributes independent variation. Given these diagnostics and our prior variable screening from the correlation heat map, we retain one representative from any highly related set and proceed with the reported specification. As a check, dropping the highest-VIF covariates one at a time yields qualitatively similar estimates for our main effects.

**Supplementary Note 3:** Detailed algorithm of the hierarchical probabilistic conformal prediction model

## 3.1 Modelling outage occurrences via point processes

Define $T$ as the time horizon of the outage trajectory, $\mathcal{S} \subseteq \mathbb{Z}_+$ as the substation index set, and $\mathcal{M} = \mathbb{Z}_+ \times \mathbb{R}$ as the mark space consisting of the number of customers influenced and the outage duration. We denote an outage event by the tuple $(t, s, m) \in [0, T) \times \mathcal{S} \times \mathcal{M}$. The outage events that we use for fitting the model is denoted as

$$(t_1, s_1, m_1), (t_2, s_2, m_2), (t_3, s_3, m_3), \ldots$$

Our modelling begins by assuming the event occurrence is governed by a conditional intensity function, which describes the rate of occurrence of the event given the history observation:

$$\lambda(t, s, m \mid \mathcal{H}_t) := \lim_{\Delta t, \Delta m \to 0} \frac{\mathbb{E}[\mathbb{N}_s([t, t + \Delta t) \times [m, m + \Delta m)) \mid \mathcal{H}_t]}{\Delta t \times \Delta m},$$

where $\mathbb{N}_s(\cdot)$ is the counting measure associated with the $s$-th substation. We assume that it is parameterized in a separable form:

$$\lambda(t, s, m \mid \mathcal{H}_t) = \lambda_g(t, s \mid \mathcal{H}_t) \cdot f(m \mid t, s),$$

where the ground process $\lambda_g(t, s)$ s defined as a self-exciting Hawkes process:

$$\lambda_g(t, s) = \mu_s(t) + \int_{[t_n \times t) \times \mathcal{S} \times \mathcal{M}} \kappa(t, t') \cdot 1\{s' \in \mathcal{N}(s)\} \, d\mathbb{N}(t', s', m'),$$

where the base rate function and the pairwise kernel function are defined as

$$\mu_s(t) = \mu_s \cdot (1 + c \cdot t) + \phi(X_t), \quad \kappa(t, t') := \alpha \cdot \beta \cdot \exp(-\beta \cdot |t - t'|), \quad \forall s \in \mathcal{S}.$$

Here $\phi$ is a linear function of the covariates across all substations at time $t$, denoted as $X_t \in R^{|\mathcal{S}| \times d}$. The biases $\{\mu_s\}_{s \in \mathcal{S}}$ are fixed constants characterizing the inherent base rate across substations. The kernel function $\kappa$ is a standard exponential decaying kernel. The function $\mathcal{N}: \mathcal{S} \to 2^{\mathcal{S}}$ is a mapping that finds the neighbourhood substation index set of a given substation index $s$. We pre-specify it as the top $\kappa$-nearest substation.

The mark conditional density function $f(m \mid t, s)$ is parameterized as a one-layer neural network that takes the spatio-temporal datum $t$ and $s$ as input, yielding its final output through a softmax activation layer:

$$f(\cdot \mid t, s) = \text{Softmax}(\text{NN}_1(t, s; \theta)),$$

where $NN_1: \mathcal{S} \times \mathcal{T} \to \mathbb{R}^{|\mathcal{M}|}$ where $\theta$ denotes the parameters within $NN_1$. The entire process is fitted by maximizing the log-likelihood using stochastic gradient descent over all parameters defined above (Kingma & Ba, 2017). Specifically, the log-likelihood function is defined as

$$\ell = \int_{[0,T) \times \mathcal{S} \times \mathcal{M}} \log \lambda(t, s, m | \mathcal{H}_t) d(t, s, m) - \int_{[0,T) \times \mathcal{S} \times \mathcal{M}} \lambda(t, s, m | \mathcal{H}_t) d\mathbb{N}(t, s, m),$$

where $\ell$ is a function of the parameters $\alpha, \beta, c, \phi, \{\mu_s\}_{s \in \mathcal{S}}$ and $\theta$. The optimization process is terminated when the loss converges.

### 3.2 Uncertainty quantification via conformal calibration

After the Hawkes process model is fitted, we make predictions from the Hawkes process model using the thinning algorithm. Specifically, for any given time interval $[T_0, T_0 + \Delta T)$ and the corresponding history trajectory $\mathcal{H}_{T_0}$ provided, the thinning algorithm allows us to draw a simulated trajectory over this time interval following the estimated conditional intensity function $\lambda$. By iteratively carrying out this procedure by sliding a fixed window forward (e.g., one month as is done in this paper), and also concatenating the previously simulated trajectory to the history, we can obtain a long range of simulated trajectories:

$$\underbrace{(\hat{t}_1, \hat{s}_1, \hat{m}_1), \dots, (\hat{t}_{N_1}, \hat{s}_{N_1}, \hat{m}_{N_1})}_{\text{Simulated in } [T_0, T_0 + \Delta T)}, \underbrace{(\hat{t}_{N_1+1}, \hat{s}_{N_1+1}, \hat{m}_{N_1+1}), \dots, (\hat{t}_{N_2}, \hat{s}_{N_2}, \hat{m}_{N_2})}_{\text{Simulated in } [T_0 + \Delta T, T_0 + 2\Delta T)}, \dots.$$

Then, by aggregating the simulated trajectories spatio-temporally (i.e., scaling each event by its mark to obtain the aggregated outage count, duration, and customer influence) into SAIDI, CAIDI, and SAIFI, we have now obtained the model's predicted sequence for these outage measures over time and over each substation. Here, we specify the time windows as equal-length disjoint time intervals that uniformly partition the training time interval:

$$[0, \Delta T) \cup [\Delta T, 2\Delta T) \cup \dots \cup [(n-1)\Delta T, n\Delta T) = [0, T),$$

where $\Delta T = T/n$. Without loss of generality, we focus on only *one* of SAIDI, CAIDI, and SAIFI, and we denote the predicted value at time step $i$, substation $s$ as $\hat{y}_{i,s}$. All the prediction result can be organized in matrix form as

$$\widehat{Y} = \begin{pmatrix} \hat{y}_{1,1} & \cdots & \hat{y}_{1,|\mathcal{S}|} \\ \vdots & & \vdots \\ \hat{y}_{n,1} & \cdots & \hat{y}_{n,|\mathcal{S}|} \end{pmatrix} \in \mathbb{R}^{n \times |\mathcal{S}|}$$

The above procedure can be repeated multiple times to obtain trials of predicted sequences, which we stack to an $n \times |\mathcal{S}| \times K$ matrix.

Then, we calibrate the model to obtain an estimate for its uncertainty, which will later be used to construct the prediction interval. Specifically, we consider the residual (a.k.a., nonconformity score) as

$$r_{i,s} = \min_{k=1,\ldots,K} \left| \hat{y}_{i,s}^{(k)} - y_{i,s} \right|, \qquad i = 1, \ldots, n$$

### *3.3 Constructing prediction intervals*

Define $\hat{Q}_s$ as the empirical quantile function with the residuals $\{r_{i,s}\}_{i=1}^n$, then we the prediction interval for substation $s$ for any step $i' > n$ can be defined as

$$\text{PI}_{i',s} := \left[ \min_{k=1,\ldots,K} \hat{y}_{i',s}^{(k)} - \hat{Q}_s(1-\alpha), \max_{k=1,\ldots,K} \hat{y}_{i',s}^{(k)} + \hat{Q}_s(1-\alpha) \right]$$

where $\{\hat{y}_{i',s}^{(k)}\}_{k=1}^K$ are $K$ aggregated simulations drawn from the Hawkes process model at time step $i'$ and location. This concludes the entire algorithm.

### *3.4 Covariates handling*

We construct the covariates dataset used by the prediction model as follows. There are two raw datasets that we begin with:
1. The first dataset is the covariate dataset associated with each substation, describing the substation conditions, including: Tree coverage (yearly), number of poles (yearly), and operation and maintenance spending (quarterly).
   The data spans from January 2014 to January 2024, where we augment the future periods until January 2040 with predictions interpolated by linear regression models. The spatial resolution is substation-level, and the temporal resolution is at a monthly resolution.
2. The second dataset is the weather dataset, containing weather observation data including: precipitation (in millimetres), maximum and minimum temperature (in Celsius degrees), and wind speed (in miles per hour). The data spans from January 2011 to January 2110, with the future period (under RCP 4.5 scenario, CESM-CAM5 Model) being prediction results provided by Byun & Hamlet (2025). The spatial resolution is substation-level, and the temporal resolution is at a daily level. We transform the data into a monthly level by merging them into the average value of each attribute.

The two datasets are then concatenated together to form a dataset spanning from January 2014 to January 2024 at a monthly resolution, with spatial resolution at the substation-level. It is then normalized to be used as the prediction covariate.

### *3.5 Thinning Algorithm*

The thinning algorithm (Ogata, 2006) is a rejection-sampling-based algorithm for simulating from point process models. In our implementation, we adopt an efficient variant of the original thinning algorithm, as described in Algorithm 1.

**Algorithm 1** Joint thinning algorithm for simulating point process

Require: Parameters: $\theta$, historic data $\mathcal{H}_0$, forecast horizon $\Delta T$, mark space $\mathcal{I}$;
1. **Initialize** $\mathcal{H}_t = \mathcal{H}_\mathcal{T}, t = 0, i \sim uniform(\mathcal{I})$;
2. **while** $t < \Delta T$ **do**
3.     Sample $i' \sim$ uniform($\mathcal{I}$); $u \sim$ uniform(0,1); $D \sim$ uniform(0,1);
4.     $x' \leftarrow (t, i'); \bar{\lambda} \leftarrow \lambda(x'|\mathcal{H}_t)$;
5.     $t \leftarrow t - \log u / \bar{\lambda}$;
6.     $x \leftarrow (t, i); \tilde{\lambda} \leftarrow \lambda(x|\mathcal{H}_t)$;
7.     **if** $D\bar{\lambda} > \tilde{\lambda}$ **then**
8.         $\mathcal{H}_t \leftarrow \mathcal{H}_t \cup \{(t, i)\}; m' \leftarrow m$;
9.     **end if**
10. **end while**
11. **return** A set of continued simulated events $\mathcal{H}_{\Delta\mathcal{T}}$, ordered by time.

**Supplementary Note 4:** Baseline comparison experiment

In this part, we show that our adopted prediction method enjoys superior accuracy by conducting a simple comparative numerical experiment. We compare against a naive vector autoregression (VAR) model (Stock & Watson, 2001), which is a commonly used time series model for multivariate forecasting tasks similar to our setting. We only consider specifying VAR with one-month and two-month lag configurations, as we assume that the outage influence is short-term. The experiment is an out-of-sample prediction task, where both models have access to 1 year of data for training, and then leave the remaining 9 years of data as test data for evaluation.

The evaluation results are reported as follows. Each entry shows the mean squared error (MSE) averaged across all spatial units and time steps, with the standard error corresponding to the variability of substation-wise MSEs over time. Across all resilience metrics, our adopted model achieves competitive accuracy, attaining the best performance on SAIDI and CAIDI (bolded), and near-optimal performance on SAIFI. These results highlight its strength in modelling outage occurrences and support our choice of adopting this approach.

| Method | SAIDI | SAIFI | CAIDI |
|---|---|---|---|
| VAR (1 month lag) | $2.39 \times 10^{-5} \pm 5.00 \times 10^{-5}$ | $\mathbf{6.24 \times 10^{-6} \pm 5.66 \times 10^{-6}}$ | $4.65 \times 10^2 \pm 9.88 \times 10^2$ |
| VAR (2 months lag) | $2.65 \times 10^{-5} \pm 3.59 \times 10^{-5}$ | $6.31 \times 10^{-6} \pm 5.66 \times 10^{-6}$ | $5.51 \times 10^2 \pm 1.08 \times 10^3$ |
| Our adopted approach (Zhou et al., 2024) | $\mathbf{1.99 \times 10^{-5} \pm 2.89 \times 10^{-5}}$ | $6.46 \times 10^{-6} \pm 5.11 \times 10^{-6}$ | $\mathbf{4.29 \times 10^2 \pm 9.42 \times 10^2}$ |

Prediction accuracy comparison results between our adopted forecast approach and the vector autoregression model with two lag configurations.

**References for supplementary information:**